\documentclass[aps,prb,a4paper,twocolumn,superscriptaddress,floatfix,nofootinbib]{revtex4-1}

\pdfoutput=1

\usepackage{graphicx} % to include figure files
\usepackage{amsmath} 
\usepackage{array} 
\usepackage{amssymb} 
\usepackage{amsfonts} 
\usepackage{color} 
\usepackage{enumitem} 
\usepackage{bm} 
\usepackage{verbatim} 
\usepackage{hyperref} 
\usepackage{morefloats} 
\usepackage{dcolumn} 
\usepackage{xspace}
\usepackage{relsize} 
\usepackage{footnote} 
\usepackage{lipsum}

\hypersetup{ citecolor=blue, colorlinks=true, urlcolor=blue }

\newcommand{\vektor}[1]{{\bf #1}}

\newcommand{\la}{\langle} 
\newcommand{\ra}{\rangle}

\newcommand{\be}{\begin{equation*}} 
\newcommand{\ee}{\end{equation*}}
\newcommand{\bea}{\begin{eqnarray*}}
\newcommand{\eea}{\end{eqnarray*}} 
\newcommand{\bra}[1]{\langle #1|}
\newcommand{\ket}[1]{|#1\rangle}

\newcommand{\eref}[1]{Eq.~(\ref{#1})}
\newcommand{\fref}[1]{Fig.~\ref{#1}}
\newcommand{\sref}[1]{Sec.~\ref{#1}}

\newcommand{\etal}{ \emph{et al.}}

\makeatletter \def\blfootnote{\xdef\@thefnmark{}\@footnotetext} \makeatother

\begin{document}

\title{Phase diagram of the quantum Ising model with long-range interactions on an
infinite-cylinder triangular lattice} 

\author{S.~N.~Saadatmand} 
\email{n.saadatmand@griffith.edu.au}
\affiliation{Centre for Quantum Computation and Communication Technology (Australian Research Council), 
Centre for Quantum Dynamics, Griffith University, Brisbane, Queensland 4111, Australia}
\affiliation{ARC Centre of Excellence for Engineered Quantum Systems,
  School of Mathematics and Physics, The University of Queensland, St
  Lucia, QLD 4072, Australia}
\author{S.~D.~Bartlett}
\affiliation{ARC Centre of Excellence for Engineered Quantum Systems,
  School of Physics, The University of Sydney, Sydney, NSW 2006,
  Australia}
\author{I.~P.~McCulloch}
\affiliation{ARC Centre of Excellence for Engineered Quantum Systems,
  School of Mathematics and Physics, The University of Queensland, St
  Lucia, QLD 4072, Australia}

\begin{abstract}
% will be added at the very end.

  Obtaining quantitative ground-state behavior for geometrically-frustrated quantum magnets with long-range interactions is 
  challenging for numerical methods. Here, we demonstrate that the ground states of these systems on two-dimensional lattices can be efficiently obtained
  using state-of-the-art translation-invariant variants of matrix product states and 
  density-matrix renormalization-group algorithms. We use these methods to calculate the fully-quantitative ground-state
  phase diagram of the long-range interacting triangular Ising model with a transverse field on 6-leg infinite-length 
  cylinders, and scrutinize the properties of the detected phases. We compare these results with those of the corresponding nearest neighbor model. 
  Our results suggest that, for such long-range Hamiltonians,
  the long-range quantum fluctuations always lead to long-range correlations, where correlators exhibit power-law decays instead of
  the conventional exponential drops observed for short-range correlated gapped phases. Our results are relevant for comparisons with recent ion-trap
  quantum simulator experiments %~[Britton\etal, Nature 484, pp.~489-492 (2012), Bohnet\etal, Science 352, 6291, pp.~1297-1301 (2016)], 
  that demonstrate highly-controllable long-range spin couplings for several hundred ions.
  
\end{abstract}

\date{\today}

\maketitle

\section{Introduction}
\label{sec:intro}

%%% Paragraph 1: a bit of history on SR and LR interations
The zero-temperature physics of geometrically-frustrated magnets with short-range (SR) 
interactions, i.e., interactions decaying exponentially with distance, is relatively 
well understood~\citep{Diep04_book,Lhuillier05,Schollwock10_book,Lacroix11_book,Sachdev11_book,Farnell14}. 
A frustration-free spin system with dominant antiferromagnetic (AFM) local couplings commonly exhibit a 
bipartite N\'eel-type~\citep{Neel48_original} ground state; while in Heisenberg-type models,
frustration can lead to the stabilization of a variety of 
exotic forms of the quantum matter such as spin glasses~\citep{Diep04_book,Lacroix11_book,Sachdev11_book},
topological~\citep{Wen02,Wen07_book,Chen10} and algebraic~\citep{Wen02} 
spin liquids, and many-sublattice long-range 
order~\citep{Lhuillier05,Schollwock10_book,Diep04_book,Lacroix11_book,Farnell14,Sachdev11_book}.
In contrast, little is known about the properties of long-range (LR) interacting spin
systems, with or without frustration, in particular for lattice dimension greater than one. (For results on 
LR-interacting AFM Heisenberg-type chains, see Refs.~\onlinecite{Sandvik10,Gong16a,Gong16b,Fey16,Jaschke17,Manmana17} and also below.) 
In this context, LR refers to  interactions decaying as 
$1/r^\alpha$, where $r$ denotes the real-space distance between 
two sites measured in units of the lattice spacing. %(and $\alpha>1$ if the thermodynamic limit is considered -- see below).
For  example, $\alpha=2$ corresponds to natural monopole-dipole-type interactions, and $\alpha=3$ to 
dipole-dipole-type atomic couplings.  We do not yet have a complete theory that would govern the physics of such 
LR-interacting Hamiltonians in two dimensions.  
In particular, consider the LR-interacting triangular quantum Ising model (defined in details below).  Due to %aforementioned obstacles
its two-dimensional arrangement, high degree of geometrical frustration, and the long-range nature of the couplings,
the ground state properties of this system are not yet fully understood.  

%%% Paragraph 1-2:
Recently, the LR-interacting triangular quantum Ising model has been simulated \emph{experimentally} with ions confined in a Penning trap~\citep{Britton12,Bohnet16} (see also Refs.~\onlinecite{Labuhn16,Garttner12}).  These experiments simulate LR interactions of hundreds of spins on a two-dimensional lattice, and it is believed that classical numerical simulations for generic LR Hamiltonians on systems of this size will be intractable.  This perceived classical intractability is a principal motivation for the development of ``quantum simulations''~\cite{Feynmann82,Lloyd96}.  Experiments that implement quantum simulations can efficiently access the physics of quantum many-body systems, whereas exact classical simulations would have a complexity that scales 
exponentially with the number of spins. (See Refs.~\onlinecite{Buluta09,Hauke12} for 
reviews and critical discussions of engineered quantum simulators.) 
%Potentially, a quantum simulator can solve a non-P complex minimization problem with a polynomial cost. 
% In the nick of time, many quantum simulators have been built and used to simulate quantum magnets by achieving 
% high level of control in trapped, ultra-cold neutral atoms or ions.
% in particular, Refs.~\onlinecite{Britton12,Bohnet16} achieved the emulation of 
% a frustrated magnet with LR Ising-type interactions with a record number of few hundred spins (see also below for details).

%%% Paragraph 1-3:
In this paper, we demonstrate that modern well-controlled approximate numerical methods can be used to probe this regime.  Specifically, we establish that
state-of-the-art variants of translation-invariant matrix product states 
(MPS)~\citep{AKLT87_original,Perez-Garcia07,McCulloch07,McCulloch08,Michel10,Schollwock11,Orus14} and density-matrix renormalization-group
(DMRG)~\citep{White92_original,White93_original,McCulloch07,McCulloch08,Schollwock11} can be used to find the 
detailed phase diagram of the LR-interacting triangular quantum Ising Hamiltonian on infinite cylinders.  These results constitute an important first step in assessing whether or not the physics of LR-interacting quantum many-body systems, now accessible through quantum simulator experiments~\citep{Britton12,Bohnet16,Labuhn16}, can also be accessed through classical numerical simulation methods.  Our results give strong evidence that they can be. Furthermore, 
we note that our results of the LR-interacting triangular quantum Ising model on cylinders 
are (to the best of our knowledge) the first attempt to create an infinite-size MPS/DMRG phase diagram of any two-dimensional LR-interacting model.

\subsection{Characteristics of LR-interacting quantum magnets}

%%% Paragraph 2: anamolies in LR-interacting systems
Long-range interacting spin systems exhibit some peculiar characteristics in comparison to their 
short-range interacting counterparts. Most strikingly, the presence of long-range interactions can break continuous 
symmetries in low-dimensions\cite{Maghrebi17}, which is strictly forbidden for SR-interacting
Hamiltonians due to the Mermin-Wigner-Hohenberg theorem~\citep{MerminWagner66_original,Hohenberg67_original}. Examples of 
symmetry-breaking due to long-range interactions include the XXZ chain exhibiting $U(1)$-symmetry-breaking at  
zero temperatures~\citep{Maghrebi17} and the square-lattice XXZ model exhibiting $U(1)$-symmetry-breaking at finite temperatures~\citep{Peter12}. 

%%% Paragraph 2-1:
Furthermore, while SR-correlated gapped phases in low-dimensions 
collectively obey an area-law for the entanglement entropy~\citep{Hamma05_PRA,Hamma05_PLA,KitaevPreskill06_original,Levin06_original}, 
Koffel\etal~\citep{Koffel12} suggests the existence of sub-logarithmic corrections to, or 
the breakdown of, the area law in LR-correlated states for $\alpha<2$. %(we revisit this claim later in the paper). 
Gong\etal~\citep{Gong17} has recently established that, for arbitrary-dimension LR-interacting systems, a `dynamical' 
variant of the area law holds for $\alpha > \text{Dim}+1$, considering the rate of entanglement entropy growth of time-evolved states (see also Ref.~\onlinecite{Schachenmayer13}),
and $\alpha > 2(\text{Dim}+1)$, considering the entanglement entropy of the ground states of an effective Hamiltonian.

%%% Paragraph 2-2:
For the purpose of the current study, the most relevant distinction between SR and LR interactions emerges from 
the realization~\citep{Vodola16} that, for the LR-interacting transverse-field Ising chain, the paramagnet and 
$Z_2$-symmetry-broken AFM ground states exhibit a bulk spin gap
(spin-flop excitations) and, although the correlations drop exponentially for short distances, %but, importantly, 
the decay is algebraic (power-law) for long distances. We contrast this behavior with the
nearest neighbor Ising model, which exhibits short-range correlated paramagnetic and AFM groundstates, and 
where power-law correlations occur only at the second-order transition in between these two phases.
Moreover, in the square-lattice XXZ model with dipole-dipole LR interactions, the 
Ising-type AFM ground state also exhibits power-law-decaying correlation functions~\citep{Peter12}. Such power-law decays are distinct
from the exponential-decaying area-law-obeying correlations observed in SR-correlated phases.
% In the following, we too provide a robust numerical proof for that 
% highly-frustrated LR-interacting systems (i.e.~the triangular lattice on infinite-length cylinders), the LR quantum fluctuations
% naturally lead to slow algebraically-decaying correlation functions.

\subsection{Details of the LR Ising Hamiltonian}

%%% Paragraph 3: details of the experimentally-engineered Hamiltonian
%Let us now review the details of the Hamiltonian that was engineered in experiments by Britton\etal~\citep{Britton12} and Bohnet\etal~\citep{Bohnet16}. 
%succeeded to demonstrate a LR-interacting Ising-type quantum simulator in two dimensions. 
The specific Hamiltonian that will be the focus of our investigation is the antiferromagnetic LR-interacting triangular quantum Ising model (LR-TQIM) 
with a transverse field.  It can be written as
\begin{equation}
\label{eq:HamLR}
  H_{LR} = J \sum_{i>j} \frac{1}{r_{ij}^\alpha} S^z_i S^z_j + \Gamma \sum_{i} S^x_i,
\end{equation}
where $i$ and $j$ specify physical sites on vertices of the triangular lattice, $r_{ij}$ denotes the real-space (chord) distance between site $i$ and $j$, and we set $J=1$ as the unit of energy. 
For $\alpha\rightarrow\infty$, $H_{LR}$ reduces to the nearest neighbor model (NN-TQIM),
\begin{equation}
\label{eq:HamNN}
  H_{NN} = \sum_{<i,j>} S^z_i S^z_j + \Gamma \sum_{i} S^x_i,
\end{equation}
where $<i,j>$ stands for summing over only NN spins.  The low-temperature properties of this NN model
are generally well-understood (see~Refs.~\onlinecite{Penson79,MoessnerSondhi01,MoessnerChandra01,Isakov03,Powalski13} and also below).

%%% Paragraph 3-2:
The experiments by Britton\etal~\citep{Britton12} and Bohnet\etal~\citep{Bohnet16} engineered 
a variable-range many-body model of hundreds of LR-interacting spin-$\frac{1}{2}$ $^9$Be$^+$ ions on a triangular lattice, 
using a disk-shaped Penning trap with single-spin readout capability.
These experiments established 
that it is practical to construct the Hamiltonian of \eref{eq:HamLR} for hundreds of spins, in the regime of $0\leq\alpha\leq3.0$. 
For such a finite set of spins and vanishing $\Gamma$ (the classical model), Britton\etal~\citep{Britton12} observed power-law decay of 
spin correlations for a variety of $\alpha$-values. Moreover, they verified the existence of a 
power-law-decaying AFM ground state for $0.05 \lesssim \alpha \lesssim 1.4$ using a mean-field theory approach.  
%Furthermore, such experimental setup allows for direct  measurement of local magnetizations through a single-spin-resolving imaging system.

%%% Paragraph 3-3:
Many possibilities for further research are opened up by these experiments, such as the 
possibility of experimental simulations of spin dynamics in two dimensions, and
effects of disorder and many-body localization, e.g.,~see~\onlinecite{Burin15}. 
Although energy scaling arguments\cite{Alet17} suggest that many-body localization does \emph{not}
occur in $\text{Dim}>1$, at least in the thermodynamic limit, signatures of localization
have been observed in two-dimensional disordered optical lattices\cite{Choi16}.
Localization has also been observed in small ion trap systems of up to 10 long-range interacting
spins\cite{Smith16,Xu18}. Penning traps offer an order of magnitude increase in the number of spins,
which makes them an ideal setup for simulating two dimensional physics.

\subsection{Existing results on the nature of LR-TQIM}

%%% Paragraph 4: review of existing theoretical works
Previous analytical and numerical works on LR-TQIM and its NN limit have provided some preliminary understanding of the physics.
For the classical NN model (i.e.~$\alpha\rightarrow\infty$, as in \eref{eq:HamNN}, 
and in the absence of the field, $\Gamma=0$), thermodynamic-limit historical studies exist: 
the lowest-energy state is a highly (macroscopic) degenerate finite-entropy phase at all finite temperatures~\citep{Wannier50_original};
this phase exhibits \emph{no} long-range order, $T=0$ being the N\'eel critical point~\citep{Stephenson70}, 
while the ground state exhibits critical $\la S^z_0 S^z_r \ra$ correlations decaying oscillatory as $\frac{1}{r^{1/2}}$. 
For finite values of the field in \eref{eq:HamNN}, using quantum-to-classical Suzuki mapping~\citep{Suzuki76},  
we note that NN-TQIM corresponds to a finite-temperature classical ferromagntically stacked layers of triangular  
AFM Ising planes (effectively replacing $\Gamma$ with the temperature for the classical 3D model). The latter system also has 
as macroscopically degenerate ground state (however, without the finite entropy). Interestingly, it 
undergoes the classical version of the ``order from disorder'' phenomenon~\citep{Penson79,Blankschtein84,Moessner00} (induced 
by thermal fluctuations), which chooses an ordered state with the expected 
wave vector of $Q^{\text{classical}}_{\text{finite-T}}=(\pm\frac{2\pi}{\sqrt{3}},\pm\frac{2\pi}{3})$ in our notation 
[i.e.~the family of three-sublattice orders that form a regular-hexagonal-shaped first Brillouin zone -- see 
below for our notation of lattice vectors]. Consistent with this, for $\Gamma\lesssim0.705$ (using our Hamiltonian conventions), 
Penson\etal~\citep{Penson79} observed the same $Q^{\text{classical}}_{\text{finite-T}}$-ordered
ground state for $H_{NN}$ with power-law-decaying correlations; above the $\Gamma_c\approx0.705$ critical point, the authors argue for another
power-law-decaying ground state with a different exponent, a finite bulk gap, and no degeneracy (we expect 
this to be the partially $x$-polarized FM phase as found below). Subsequently, 
quantum Monte Carlo (QMC) calculations~\citep{MoessnerSondhi01,Isakov03} verified the stabilization of a three-sublattice
AFM ordered ground state for the weak fields. Importantly, these authors noted that the small-$\Gamma$ NN-TQIM can be also 
mapped to a quantum dimer model ($\frac{v}{t}\rightarrow0$ limit of Rokhsar-Kivelson Hamiltonian, $H_{QDM}$) 
on a dual kagom\'{e} lattice formed by the centers of
the triangular plaquettes. Such dimer arrangements can be labeled using the so-called `height 
configurations'~\citep{MoessnerSondhi01,MoessnerChandra01}. In fact, the existence of the map to the height model already means that
the classical model should exhibit power-law correlations~\citep{MoessnerSondhi01} under a set of `reasonable' assumptions. $H_{QDM}$, 
which corresponds to the NN-TQIM, exhibits a series of uniform ground states with valence-bond solid ordering 
that translates to three-sublattice orders on the triangular lattice
(will be stated as $(S^z_a, S^z_b, S^z_c)$, whereby $S^z_{\{a,b,c\}}$ stands for local spin polarization in spins' $z$-direction of vertices $\{a,b,c\}$ for 
a three-site unit-cell formed by a triangular plaquette). Such ordering was previously
observed in Landau-Ginzburg-Wilson analyses~\citep{Alexander80,Blankschtein84} of three-dimensional 
FM stacked triangular AFM Ising lattices, where it called a 
`clock' order due to appearance of a sixfold clock term breaking the $XY$-symmetry. 
Let us now summarize the existing results on the phase diagram of NN-TQIM at $T=0$: 
Refs.~\onlinecite{Isakov03,Blankschtein84} found that the model
undergoes a quantum phase transition in the universality class of 3D-$XY$, namely from a clock order in the low fields to
a $S_z$-magnetization-disordered ($x$-polarized FM) ground state in the large fields [the existence of such universality class was later
advocated for the LR model too by Humeniuk~\citep{Humeniuk16}]. Furthermore, QMC simulations of Refs.~\onlinecite{MoessnerSondhi01,Isakov03} suggest
the selection of $(0.5, -0.5, 0)$-ordering for the clock phase having zero net 
magnetization (which corresponds to a `hierarchical' plaquette order on the dimer model). 

%%% Paragraph 4-3:
In contrast to the NN model, there are few existing studies of the long range model.  
The most comprehensive is by Humeniuk~\citep{Humeniuk16}, which presents both thermodynamic-limit mean-field analyses (for a wide range of $\alpha$)
and stochastic-series-expansion QMC simulations (only for $\alpha=3.0$) of \eref{eq:HamLR} on 
disk-shaped, open-boundary-conditioned triangular lattices hosting up to 301 spins
for a variety of $\Gamma$-values.  A semi-quantitative phase diagram is constructed 
for the model with the main high-precision QMC results only available at $\alpha=3.0$ but for a wide range of field values.
In this study, it was found that, for large enough $\alpha$, the clock-ordered 
phase chooses the sublattice structure of $(0.5, -0.25, -0.25)$, 
i.e.~the so-called $120^\circ$ order.  This result differs to the phase diagram 
of the NN model from Refs.~\onlinecite{MoessnerSondhi01,Isakov03}, and with our results (see below). 
While the $(0.5, -0.5, 0)$-ordering for the large-$\alpha$ limit is argued by the present and two other numerical studies,
we note that such a difference might be still due to the restricted lattice geometry employed in our calculations and
different handling of the QMC's inherent sign problem for the frustrated systems in Refs.~\onlinecite{MoessnerSondhi01,Isakov03} and Ref.~\onlinecite{Humeniuk16}.
Nevertheless, the quantitative phase diagram and the realization of three distinct phase regions (including the clock 
and $x$-polarized FM ordering) in Ref.~\onlinecite{Humeniuk16} is in line with our findings.

\subsection{MPS and DMRG algorithms for LR interactions}

%%% Paragraph 5: An introductory paragraph on the employed methods
Variants of MPS and DMRG algorithms (see~\onlinecite{Schollwock11,Stoudenmire12} for reviews) have already revolutionized our understanding of 
the low-energy physics of low-dimensional local Hamiltonians by providing an efficient platform for numerical simulations.
The success of these algorithms in capturing the properties of such Hamiltonians 
can be best understood through the MPS description, i.e.,~the wavefunction ansatz that underlies DMRG.  
However, when one considers LR-interacting models,
finite-size numerical approaches suffer from
the explicit existence of a cutoff or other ways of limiting the range of LR couplings and, therefore, exhibit strong boundary effects. 
As such, many of these algorithms may not capture the essential physics
associated with LR fluctuations. Later in this work, we will see some discrepancies between finite-size calculations and our infinite-size results. 

%%% Paragraph 5-2:
However, MPS algorithms that act directly in the thermodynamic limit such as iDMRG~\citep{McCulloch08}, which contain fixed-pint transfer matrix 
equations and naturally live in the thermodynamic limit at least in one spatial direction, can be more efficient for LR
models such as Eq.~\eqref{eq:HamLR}. The key innovation here was in the realization that MPS can also describe the low-energy sector
of Hamiltonians with rapidly (i.e.~\emph{exponentially}) decaying interactions~\citep{McCulloch08,Crosswhite08}. 
Specifically,
Refs.~\onlinecite{McCulloch08,Crosswhite08} established that matrix product 
operators~\citep{McCulloch07,McCulloch08,Schollwock11,Hubig17} (MPOs), which are MPS-based representations of operators, can be written  to 
include exponential-decaying couplings such $e^{-\lambda r}$ (see below). Fortunately, this method is sufficient to describe LR decays as well, since one can expand an algebraically-decaying function 
%(in the range of $r\epsilon[0,\infty)$) 
in terms of the sum of some exponential terms. As an example,
\begin{equation}
\label{eq:ExpExpansion}
  \frac{1}{r^\alpha} \simeq \sum_{i=1}^{n_{\rm cutoff}} a_i e^{-\lambda_i r}, 
\end{equation}
where $a_i$ and $\lambda_i$ are constants to be fitted, for example by using a non-linear least-square 
method. Obviously, the expansion is only exact if $n_{\rm cutoff}\rightarrow\infty$. 
%(in that case, the semi-equality in \eref{eq:ExpExpansion} can be replaced with an equality symbol).
The existence of $n_{\rm cutoff}$ means that we still face a distance scale cutoff, that is, iMPS should 
only be considered as an improvement over finite-size calculations with a fixed cutoff for the interaction lengths.
However, in practice, a small $n_{\rm cutoff}$ can often be chosen for iDMRG simulations in a way that describes the LR physics very well.  We can therefore replace a LR Hamiltonian such as $H_{LR}$  with an approximate one, $H_{\text{LR-approx}}$, in the form of
\begin{align}
\label{eq:ExpExpansionHam}
  H_{LR} &\longleftrightarrow \notag \\ 
  H_{\text{LR-approx}} &= \sum_{i>j} \Big( \sum_{k=1}^{n_{\rm cutoff}} a_k(i,j) e^{-\lambda_k r_{ij}}  \Big) S^{z}_i S^{z}_j \notag \\ 
  &+ \Gamma \sum_{i} S^x_i.
\end{align}
%

%%% Paragraph 5-2:
Consider a rather simple LR-interacting system: the one-dimensional 
exactly-solvable Haldane-Shastry model~\citep{Haldane88_original,Shastry88_original}
$H_{\text{Haldane-Shastry}} = J \sum_{i>j} \frac{1}{r_{i,j}^2} \vektor{S}_i \cdot \vektor{S}_j$.  This model
has known ground state energy per site of $-\frac{\pi^2}{24}$ in the thermodynamic limit (in the units of $J$).  A quick iDMRG calculation with $n_{\rm cutoff}=5$ (not detailed here) 
reproduces the excellent residual energy per site of $\Delta E=E_{\text{iDMRG}}-E_{\text{exact}}=1.15(2)\times10^{-5}$ for just 
$m_{\max}=100$, while Ref.~\onlinecite{Crosswhite08} kept up to 
$n_{\rm cutoff}=9$ and reproduced an energy per site with the the best accuracy of $\Delta E \approx 2\times10^{-6}$ for just number of states $m_{max}\approx200$. 
We note that having a finite $n_{\rm cutoff}$ essentially 
means all measurements on $H_{\text{LR-approx}}$ shall depend on an effective cutoff length (an effective range), namely 
$\mathcal{L}_{\rm cutoff}(n_{\rm cutoff})$ (which, in principle, also depends to the system geometry and Hamiltonian control parameters), where  
$\mathcal{L}_{\rm cutoff}(\infty)\rightarrow\infty$ reproduces the true thermodynamic limit.
In other words, the power-law decay, appearing in \eref{eq:ExpExpansion} and (\ref{eq:ExpExpansionHam}), would be almost exactly equivalent to 
the sum of exponential decays up to this $\mathcal{L}_{\rm cutoff}$, while for longer 
distances (although $H_{\text{LR-approx}}$ may provide some insights on the physics of $H_{LR}$), the former now drops
significantly faster. In fact, the cost incurred by our approximation, \eref{eq:ExpExpansionHam}, is the
introduction of a new size-dependent quantity, $\mathcal{L}_{\rm cutoff}$ (or $n_{\rm cutoff}$), where in principle one should also perform
finite size scaling over $\mathcal{L}_{\rm cutoff}$ to find the observables in the thermodynamic limit. However, 
our studies show that the relative changes on physical observables of our interest (other than the correlation lengths -- see below) are negligible for 
both $H_{\text{Haldane-Shastry}}$ (on an infinite chain) and $H_{LR}$ (on a 6-leg infinite-length cylinder), when
the number of kept exponentials is as large as $n_{\rm cutoff}=10$; this value of cutoff reproduces an
effective range of \emph{two hundred} of lattice spacings or better, $\text{min}[\mathcal{L}_{\rm cutoff}]=\mathcal{O}(100)$, for both Hamiltonians.

\subsection{Summary of our findings}

%%% Paragraph 1:
We now briefly summarize our main findings.  For the nearest-neighbor Hamiltonion $H_{NN}$ of \eref{eq:HamNN}, on 
6-leg infinite-length cylinders using iDMRG with $m_{max}=250$, we find a
phase diagram that hosts a LR-correlated three-sublattice AFM $(0.5,-0.5,0)$-type clock order for 
$\Gamma\leq0.75(5)$, a trivial SR-correlated $x$-polarized FM order for 
larger $\Gamma$, and a second-order phase transition separating them.
The AFM ground state arises as the result of $Z_2$-symmetry breaking and is stabilized against the highly-degenerate classical ground state
at $\Gamma=0$ through the ``order from disorder'' phenomenon (induced by quantum fluctuations) as previously discussed. 
These results are in agreement with Refs.~\onlinecite{MoessnerSondhi01,Isakov03,Penson79} phase diagrams. 

%%% Paragraph 1-2:
For the LR-interacting Hamiltonian $H_{LR}$ of \eref{eq:HamLR}, on 6-leg infinite-length cylinders
using iDMRG to optimize $H_{\text{LR-approx}}$ with $m_{max}=500$ and $n_{cutoff}=10$, we 
find a phase diagram that exhibits three distinct
ground states: \textit{(i)} a LR-correlated two-sublattice $Z_2$-symmetry-broken AFM columnar order for low-$\alpha$ and low-$\Gamma$ (previously unknown
for the LR model), \textit{(ii)} a LR-correlated three-sublattice $Z_2$-symmetry-broken AFM $(0.5,-0.5,0)$-type clock order for large-$\alpha$ and low-$\Gamma$ (as one 
should expect from the SR-correlated version of this phase on the NN model, although some features were previously unknown
for the LR model), and \textit{(iii)} a LR-correlated $x$-polarized FM order for any large-$\Gamma$. 
%Both phases are expected to be gapped. 
Both AFM phases are expected to possess vanishing spin gaps due to existence of robust LR correlations.
The most significant difference between the detected 
ground states of the NN model and the LR model is that all phases of the latter  exhibit LR (power-law decaying) correlations,
at least for the distances comparable to their measured correlations lengths. 
It is important to note that due to higher computational difficulties, we do \emph{not} provide finite size 
scalings with the cylinder's width for this first iDMRG study of the LR-TQIM; therefore, our provided phase diagram 
is only precise for the cylindrical boundary conditioned model and not essentially in 
the true 2D limit where $\text{width}\rightarrow\infty$. Nevertheless, our results still confirm that in ladder-type two-dimensional highly-frustrated magnets, LR quantum fluctuations always 
lead to LR correlations in the ground states. These results can
provide directions for the future ion-trap experiments and offer some foundational understandings of the physics of LR-interacting systems.
In particular, corrections to the area law of entanglement entropy is expected for such two-dimensional LR-correlated phases, as observed for their 1D counterparts.

%%% Paragraph 6: reporting the structure for the rest of the paper
The remainder of this paper is organized as follows. In \sref{sec:methods}, 
we present the employed iMPS and iDMRG algorithms in further detail, covering the 
inclusion of LR interactions in MPOs. The structure of the triangular lattice on infinite-length cylinders and the map onto 
the MPS chain is explained in the same section. The calculated phase diagrams of $H_{NN}$ and $H_{LR}$ are displayed 
and extensively commented in \sref{sec:PhaseDiagramNN} and 
\sref{sec:PhaseDiagramLR} respectively, together with analyses of the properties of each detected ground state. In \sref{sec:conlusion}, 
we conclude our findings and suggest some possible future directions.

\section{Methods}
\label{sec:methods}

%%% Paragraph 1: MPS and MPO representations
The ground state of a SR-interacting Hamiltonian on an $L$-site lattice with periodic boundary 
conditions (the translation-invariant limit will be obtained when we set $L\rightarrow\infty$) can 
be generally well-approximated %or exactly written 
using the MPS ansatz:
\begin{equation}
  \text{Tr} \sum_{\{s_i\}} \mathcal{A}^{[s_1]}_1 \mathcal{A}^{[s_2]}_2 \cdots \mathcal{A}^{[s_L]}_L 
  \ket{s_1} \otimes \ket{s_2} \otimes \cdots \otimes \ket{s_L},   
\label{eq:MPS-generic}
\end{equation}
where $\mathcal{A}^{[s_i]}_i$ is an $m \times m$ matrix that encodes all local information available to the $i$th state and 
$s_i$ capture the local $d$-dimensional physical space (for example, $s_i=\{\downarrow,\uparrow\}$ and $d=2$ for spin-$1/2$ particles)
and $m$ is referred to as the bond dimension of or the number of states in the MPS. The matrices $\mathcal{A}$  satisfy an 
orthogonality condition and can be chosen to only contain purely real values 
(see Refs.~\onlinecite{McCulloch07,McCulloch08,Schollwock11} for details). Hamiltonian
operators on this $L$-site lattice can be analogously represented in the MPO form of
\begin{align}
  \sum_{\{ s_i , s^\prime_i\}} &M^{s_1 , s^\prime_1} M^{s_2 , s^\prime_2} \cdots M^{s_L , s^\prime_L} \notag \\ 
  &\times \ket{s_1}\bra{s^\prime_1} \otimes \ket{s_2}\bra{s^\prime_2} \otimes\cdots \otimes \ket{s_L}\bra{s^\prime_L} \;,
\label{eq:MPO-generic}
\end{align}
where $M^{s s^\prime}_{a a^\prime}$ can be thought of as a rank-4 tensor: $s, s^\prime \in \{ 1,2,\ldots,d \}$ and $a, a^\prime \in \{ 1,2,\ldots,\tilde{m} \}$,
where $\tilde{m}$ is the MPO bond dimension. We note that MPOs always provide an exact representation for the physical operators
(whereas MPS is only an exact representation of the state for some very special 
wave functions, commonly having a small finite $m$, or when $m\rightarrow\infty$; see Ref.~\onlinecite{Orus14} for examples).
It is convenient to regard MPOs as $\tilde{m} \times \tilde{m}$ (super-)matrices 
where the elements are local operators (matrices) acting on local 
physical spaces. For a Hamiltonian that is a sum of finite-range interacting terms, one can
write~\citep{McCulloch07,McCulloch08,Michel10} $M$-matrices in their Schur form (e.g.~see~\onlinecite{Golub12_book});
here, we choose to present all such MPOs in their upper triangular form, since that makes it easy to read off the
form of the operator from top-left to bottom-right.
%(such forms are sometimes referred to as `triangular MPOs'). 
%Furthermore, without a loss in generality, we can always choose 
%to set the top-left and bottom-right components of an open-boundary MPO to the identity matrices~\citep{McCulloch07,Michel10}, $I$.
As a clarifying example, to represent the infinite sum of local operator of form
$\hat{A} \otimes \hat{A} \otimes \cdots \otimes \hat{A} \otimes \hat{B} \otimes \hat{D} \otimes 
\hat{E} \otimes \cdots \otimes \hat{E} \otimes \hat{E}$, containing a NN two-body term, we only need 
a $3\times3$ $M$-matrix (using transposed matrices compared with the notation of Ref.~\onlinecite{McCulloch08}) as
\begin{equation}
  M = \begin{pmatrix} \hat{A} & \hat{B} & \hat{0} \\ 
                      \hat{0} & \hat{0} & \hat{D} \\ 
                      \hat{0} & \hat{0} & \hat{E} \end{pmatrix}~.
\label{eq:M-ABCDE}
\end{equation}
This can be easily extended to represent any finite-range $N$-body term (refer to Refs.~\onlinecite{McCulloch08,Saadatmand17_thesis} for more examples). 
In past, such notion of MPOs have been widely used to describe finite-range interacting Hamiltonians.  For example,
the MPO for $H_{NN}$ given by \eref{eq:HamNN} on an arbitrary-size translation-invariant lattice corresponds to
\begin{equation}
  M_{\text{Ising}} = \begin{pmatrix} I & S^z & \Gamma S^x \\ 
                                       & 0 & S^z \\ 
                                       &   & I \end{pmatrix}~,
\label{eq:MPO-NN}
\end{equation}
where we have suppressed displaying the trivial zero elements.

%%% Paragraph 1-2:
Similar Schur-form MPOs can be used to represent exponentially-decaying operators 
in the form of the long-range string-like terms~\citep{Crosswhite08}.
This in turn provides one with an ansatz capable of describing power-law decaying Hamiltonians using \eref{eq:ExpExpansion}.
The key is in filling the additional diagonal matrix elements of a Schur-form 
MPO other than those identities on the edge row and column, i.e.,~an infinite sum of string operators in the form of 
$\hat{A} \otimes \hat{A} \otimes \cdots \otimes \hat{A} \otimes \hat{B} \otimes \hat{C} \otimes 
\cdots \otimes \hat{C} \otimes \hat{D} \otimes \hat{E} \otimes \cdots \otimes \hat{E} \otimes \hat{E}$ 
(here, we set no one-body field term like $S^x$ in \eref{eq:MPO-NN}; we still assign a two-body operator set of $\hat{B}$ and $\hat{D}$, and 
most importantly, varying-range $\hat{C}$ operators)
can be written in the MPO form of
\begin{equation}
  M_{LR} = \begin{pmatrix} \hat{A} & \hat{B} & 0 \\ 
                                   & \hat{C} & \hat{D} \\ 
                                   &         & \hat{E} \end{pmatrix}~.
\label{eq:M-LR}
\end{equation}
%
%When such $M$-matrices are summed over the length lattice, they construct an exponentially-decaying term. 
To produce the LR terms of the form in \eref{eq:ExpExpansionHam}, but for simplicity on an infinite chain, we can set 
$\hat{C} = |\lambda| \hat{I}$, with $|\lambda|<1$, $\hat{B} = \hat{S}^z$ always acting on a site numbered as $i-1$, 
$\hat{D} = \lambda \hat{S}^z$ always acting on a site numbered as $j$, and placing the identity operator elsewhere. It is straightforward to 
check that the resulting string operator is an infinite sum of the form 
$\sum_{j>i} \hat{I} \otimes \hat{I} \otimes \cdots \otimes \hat{I} \otimes \hat{S}^z_{i-1} \otimes |\lambda|^{j-i} \hat{I} \otimes
\hat{I} \otimes \cdots \otimes \hat{I} \otimes \hat{S}^z_j \otimes \hat{I} \otimes \cdots \otimes \hat{I} \otimes \hat{I}$
%$... I_{-2} I_{-1} \mathcal{M}_{0} \mathcal{M}_{1} ... \mathcal{M}_{r} I_{r+1} I_{r+2} ...$ 
corresponding to the Hamiltonian term of $\sum_{j>i} \lambda^{j-i} S^z_{i-1} S^z_j = \sum_{j>i} e^{\ln|\lambda|(j-i)} S^z_{i-1} S^z_j$.
The extension of such LR string operators to infinite cylinders would involve summing over several chain-type terms, but otherwise is straightforward. 

%%% Paragraph 1-2:
Let us now list the order parameters of our interest: the order parameter for a clock order can be considered as the magnitude of
\begin{equation}
\label{eq:Oxy}
  O_{XY} = \frac{1}{N_{XY}} ( S^z_a + e^{i\frac{4\pi}{3}} S^z_b + e^{-i\frac{4\pi}{3}} S^z_c), 
\end{equation}
where $N_{XY}$ is a normalization factor, the value of which should be set according to the $S^z_{\{a,b,c\}}$-magnitudes. We note that
$O_{XY}$ is sometimes referred to as the `$XY$ order parameter'. We work on a 
translation-invariant lattice with the unit-cell size of $L_u$ and use the following three 
order parameters to fully quantify the phase diagrams of both $H_{NN}$ and $H_{LR}$. These order parameters are
the normalized total $S_x$-magnetization per site,
\begin{equation}
\label{eq:M1x}
	M_1^x = \frac{1}{L_u} \sum_{i \in \{\text{unit-cell}\}} S^x_i\,,
\end{equation}
suited to detect the single-sublattice FM ordering; the normalized total $S_z$ staggered magnetizations per site,
\begin{equation}
\label{eq:M2z}
	M_2^z = \frac{1}{L_u} \sum_{i \in \{\text{unit-cell}\}} (-1)^i S^z_i\,,
\end{equation}
suited to detect the two-sublattice AFM columnar ordering; and $XY$ order parameter per site,
\begin{equation}
\label{eq:M3z}
	M_3^z = \frac{1}{L_u \times N_{XY}} \sum_{a,b,c \in \{\text{unit-cell}\}} S^z_a + e^{i\frac{4\pi}{3}} S^z_b + e^{-i\frac{4\pi}{3}} S^z_c\,,
\end{equation}
suited to detect the three-sublattice AFM clock ordering. 

%%% Paragraph 1-3:
Returning to the iMPS construction of the model, 
after building $M_{LR}$-type MPOs for \eref{eq:HamLR} and finite-range ones for \eref{eq:HamNN}, we then optimize the corresponding MPS using
the iDMRG algorithm, such that the reduced density matrices will then satisfy fixed-point equations.  We then use the 
method of the \emph{transfer operator}, $\mathcal{T}_I$, as explained in Refs.~\onlinecite{McCulloch08,Michel10,Saadatmand17}
[we note that the original `transfer matrix' scheme was introduced for MPS in Refs.~\onlinecite{Ostlund95_original,Rommer97}], to find the energies per site
as well as the expectation values of magnetizations per site, $M^x_1$, $M^z_2$, and $M^z_3$. We note that 
on an infinite lattice the elements of $M$-matrices would diverge, however, the expectation values per site are well-defined, and 
the principal correlation length, $\xi(m)$, can be measured from the spectrum of $\mathcal{T}_I$ 
($\xi$ is always measured per Hamiltonian unit-cell size, but due to the cylindrical
form of the lattice can be thought to represent the typical long-direction size `per lattice spacing').
Moreover, to avoid the requirement of the extra normalization~\citep{Homrighausen17} needed for some $\alpha$-values (when considering 
the thermodynamic limit or studying the scaling behavior of finite-size observables), we 
confine ourselves to $\alpha > 1$, where the thermodynamic limit is well-defined without additional normalization.
Finally, we note that due to the current limitations of the iDMRG algorithm, we were 
unable to directly calculate the bulk spin gap for any of the presented ground states in this paper.

\begin{figure}
  \begin{center}
    \includegraphics[width=0.99\columnwidth]{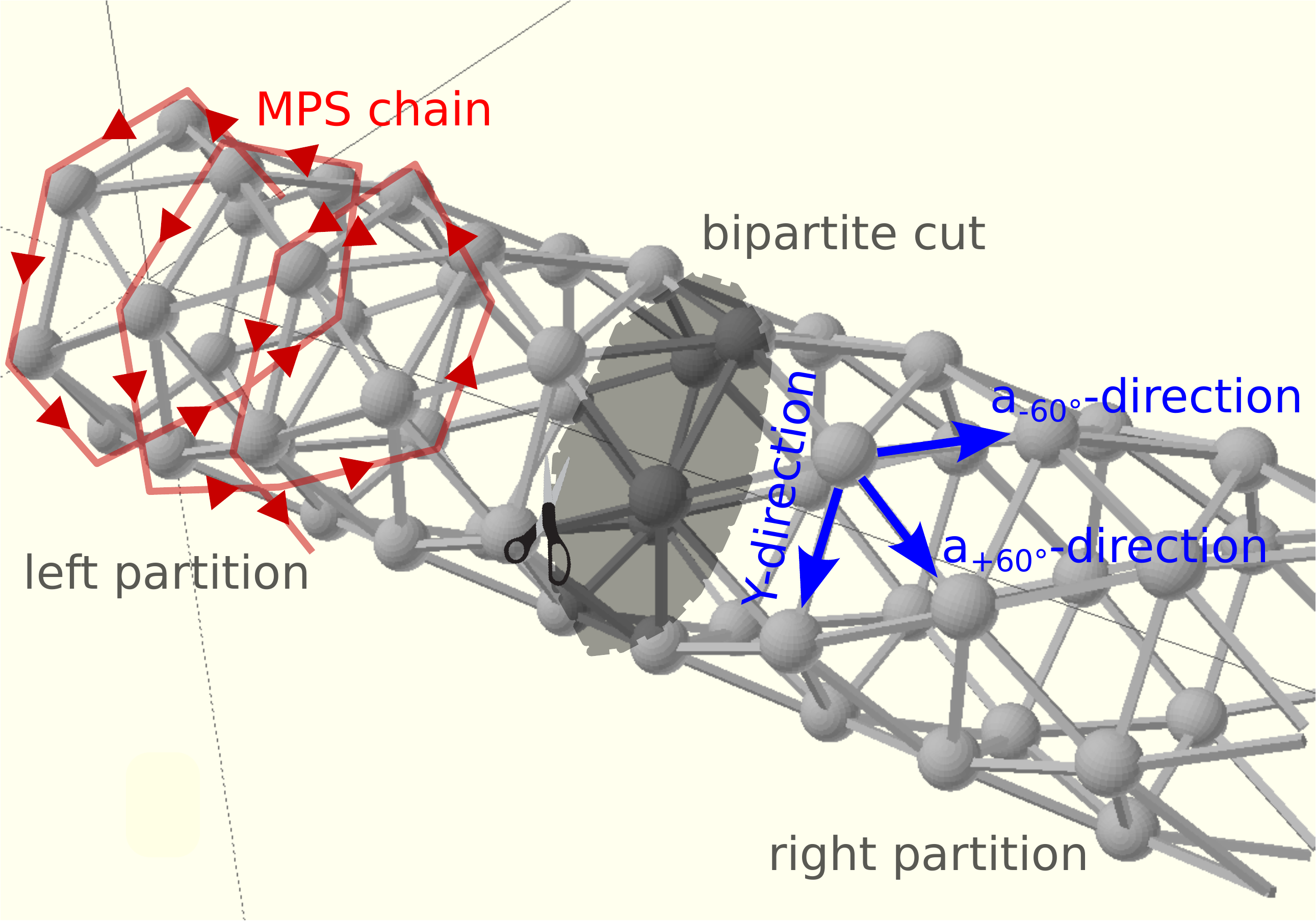}
    \caption{(Color online)
    Cartoon visualization of a triangular lattice on a YC6 cylinder. Spins
    sit on spheres. An `efficient' mapping of the MPS chain is shown using the red spiral.
    The green arrows represent the unit vectors on three principal lattice directions.
    The transparent gray plane corresponds to the bipartite cut that creates a left and right 
    partition, without crossing any $Y$-direction bond, and is used to calculate bipartite 
    iDMRG quantities. 
    \label{fig:TriYC6-generic}}
\end{center}
\end{figure}

%%% Paragraph 3: Mapping of the MPS chain and the wrappings of infinite cylinders
As is clear from \eref{eq:MPS-generic}, MPS is inherently a 1D ansatz. Therefore, DMRG simulations 
in 2D require a mapping between the MPS chain and the physical lattice.  Unavoidably, this means that interactions  
in the 2D lattice map to couplings that are at least as long range, but often longer range, on the 1D MPS chain. We use 
an `efficient' mapping for the MPS onto an infinite-length 2D lattice 
(i.e.~$L_x\rightarrow\infty$ and $L_y$ be finite, where $L_x$($L_y$) always denotes long- (short-) direction size), 
as demonstrated in Fig.~5.2(a) of Ref.~\onlinecite{Saadatmand17_thesis} and detailed in its corresponding section (see also below). This particular mapping 
minimizes the range of finite-range couplings in the resulting 1D Hamiltonian. Furthermore, there exists an infinite
number of ways to wrap a 2D lattice to create a generic periodic boundary condition in the $Y$-direction.  
The wrapping creates an infinite-length cylinder, one which is generally twisted. The use 
and classification of such cylindrical boundary conditions are common practice in the study of single-wall carbon nano-tubes (e.g.~see~\onlinecite{Wilder98}).
To identify the wrappings of the triangular lattice on an infinite cylinder, we use this standard but versatile
classification, where the corresponding notations are detailed in Chapter 2 of Ref.~\onlinecite{Saadatmand17_thesis}. 
The majority of our calculations are performed on the highly convenient and 
computationally beneficial infinite-length YC6 structures with the shortest possible wrapping vector as
$\vektor{C}_0[\text{YC}6]\!=\!(6,-6)$ in the unit of $(\vektor{a}_{+60^\circ},\vektor{a}_{-60^\circ})$ (see~\fref{fig:TriYC6-generic} -- 
in this paper, we represent the unit vectors of inverse lattices, $(\vektor{K}_{x},\vektor{K}_{y})$, in correspondence 
to the $(\vektor{a}_{+60^\circ},\vektor{a}_{-60^\circ})$ notation).
Some benefits of the YC structure include having the same circumferences 
as the short-direction size of $L_y$, iDMRG Hamiltonian unit-cell aligning in the $\vektor{C}_0$-direction, 
and providing high-resolving power for the spectrum of reduced density matrix while respecting the bipartite and tripartite lattice symmetries
with some non-excessive wave function unit-cell sizes. A generic 
YC6 structure is demonstrated in \fref{fig:TriYC6-generic}, where we also display the MPS efficient mapping method. 
We reiterate that, on the cylinder, such mapping provides the shortest one-dimensional 
SR coupling ranges over the periodic boundary condition connections. 
To study the effect of the lattice geometry on detected phases, we perform few additional iDMRG calculations on
distinctly wrapped systems, namely XC6 structures with $\vektor{C}_0[\text{XC}6]=(6,6)$ and 
6-leg three-site unit-cell structures with $\vektor{C}_0[\text{three-site}]=(6,-2)$, 
for some control parameters of interest (see below for details).

%%%%%%%%%%%%%%%%%%%%%%%%%%%%%%%%%%%%%%%%%%%%%%%%%%%%%%%%%%%%%%%%%%%%%%%%%%%
%%%%%%%%%%%%%%%%%%%%%%%%%%%%%%%%%%%%%%%%%%%%%%%%%%%%%%%%%%%%%%%%%%%%%%%%%%

%%% Paragraph 4: Details of actually employed structures and simulations
In practice, we find the phase diagram of the Hamiltonian $H_{LR}$ of \eref{eq:HamLR} mainly by performing an extensive series of ground state 
iDMRG simulations on YC6 structures for $\alpha=[1.1,4.0]$ and $\Gamma=[0.1,1.5]$, having a resolving power as small as
$\Delta\alpha,\Delta\Gamma=0.05$ and maximum MPS bond dimension of $m_{\max}=500$. We 
used a 10-term expansion ($n_{\text{cutoff}}=10$) of the form \eref{eq:ExpExpansion} 
to translate exponential decays, produced by the MPO of \eref{eq:M-LR},
into LR interactions. We reiterate that our Hamiltonian reconstruction and validation tests proved 
that a 10-term expanded $H_{\text{LR-approx}}$ of the form \eref{eq:ExpExpansionHam} can 
faithfully describe (before terms start to fall exponentially rapidly) the original Hamiltonian, $H_{LR}$, 
on the YC6 structure, typically, up to \emph{few hundreds} of lattice spacings (the exact value of 
$\mathcal{L}_{\text{cutoff}}$ depends on the assigned Hamiltonian control parameters). The selection of $L_y=6$ is mainly due to the simplicity as this is 
the smallest width for which the YC structure can be set to respect the $Y$-axis bipartite and 
the tripartite symmetries. However, we note that $L_y=6$ is large enough to produce a
phase diagram exhibiting exclusively two-dimensional phase phenomena, some of which are distinct from
the phase properties observed in 1D long-range quantum Ising model~\citep{Vodola16}. Additionally,
we remind that our width-6 results are a first attempt to create an iMPS/iDMRG phase diagram 
for this model (or any two-dimensional LR-interacting spin system). 
As mentioned, we also study the LR-TQIM on XC6 and 
$\vektor{C}_0[\text{three-site}]=(-3,3)$ systems for $(\alpha,\Gamma,m_{\max})=(1.5,0.2,100)$ [predicted
to lie deep inside the LR columnar phase region -- see below] and a series of very large $\alpha$ and small $\Gamma$ values with $m_{\max}=250$
[predicted to lie deep inside $(0.5,-0.5,0)$ clock phase region -- see below]. Our results show that
the energy per site and real-space correlation patterns are the same in comparison to the equivalent points on YC6 systems up to
the machine precision. These results confirm that the stabilization of multi-partite ground states of LR-TQIM
is independent of the geometry, i.e., the choice of the wrapping structure hosting the triangular lattice.
%therefore, we stick to presenting the results only for YC systems in the rest of the paper. 
For obtaining the phase diagram of $H_{NN}$, we perform conventional finite-range iDMRG calculations associated
with the MPO \eref{eq:MPO-NN} for 11 chosen points distributed unevenly in $\Gamma=(0,2.0]$, while keeping up 
to $m_{\max}=250$ number of states.

%%%%%%%%%%%%%%%%%%%%%%%%%%%%%%%%%%%%%%%%%%%%%%%%%%%%%%%%%%%%%%%%%%%%%%%%%%%
\begin{figure}
  \begin{center}
    \includegraphics[width=0.99\columnwidth]{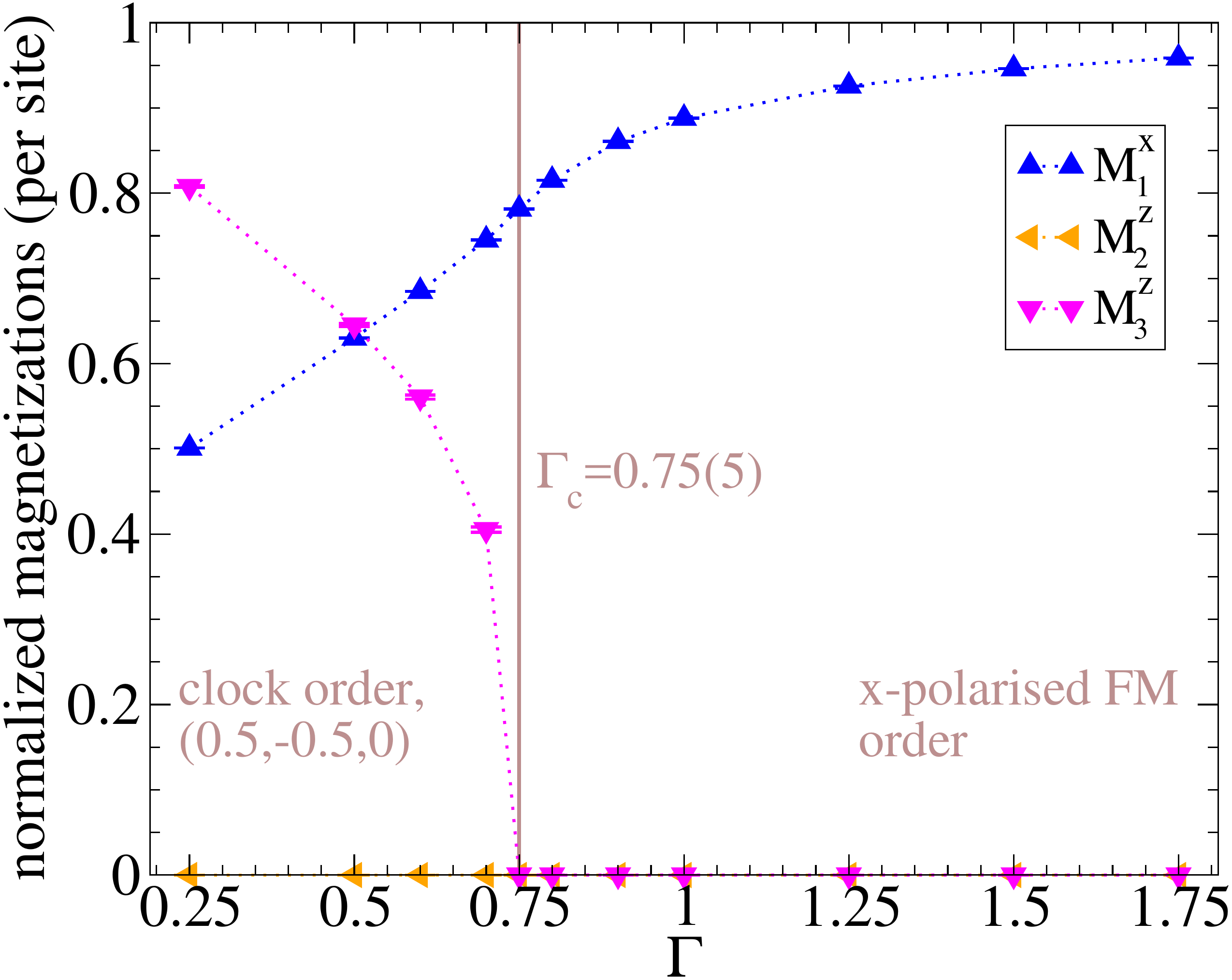}
    \caption{(Color online)
    iDMRG phase diagram of NN-TQIM, \eref{eq:HamNN}, on a YC6 structure. Filled triangular symbols
    with error bars show the thermodynamic-limit magnetizations per site, 
    $M_1^x(m\rightarrow\infty)$, $M_2^z(m\rightarrow\infty)$, and $M_3^z(m\rightarrow\infty)$ [cf.~Eqs.~\ref{eq:M1x}, \ref{eq:M2z}, and \ref{eq:M3z} respectively],
    extrapolated using a linear fit versus $\sqrt{\varepsilon_m}$ (see below for some examples on individual fits). 
    We set $N_{xy}=\frac{1}{\sqrt{12}}$ for $M_3^z$ [as appeared in \eref{eq:Oxy}], which is the maximum value achievable 
    for an ideal $(0.5,-0.5,0)$ ordering. The symbols with no error-bars stand for $M_2^z(m_{\max})$ and 
    $M_3^z(m_{\max})$, where no analytical fit was possible toward the thermodynamic
    limit of $m\rightarrow\infty$ (due to extreme decay and/or smallness of observables). Pointed 
    lines only connect symbols as guides for the eyes.
    \label{fig:PhaseDiagramNN}}
\end{center}
\end{figure}
%%%%%%%%%%%%%%%%%%%%%%%%%%%%%%%%%%%%%%%%%%%%%%%%%%%%%%%%%%%%%%%%%%%%%%%%%%

%%%%%%%%%%%%%%%%%%%%%%%%%%%%%%%%%%%%%%%%%%%%%%%%%%%%%%%%%%%%%%%%%%%%%%%%%%%
\begin{figure}
  \begin{center}
    \includegraphics[width=0.49\columnwidth]{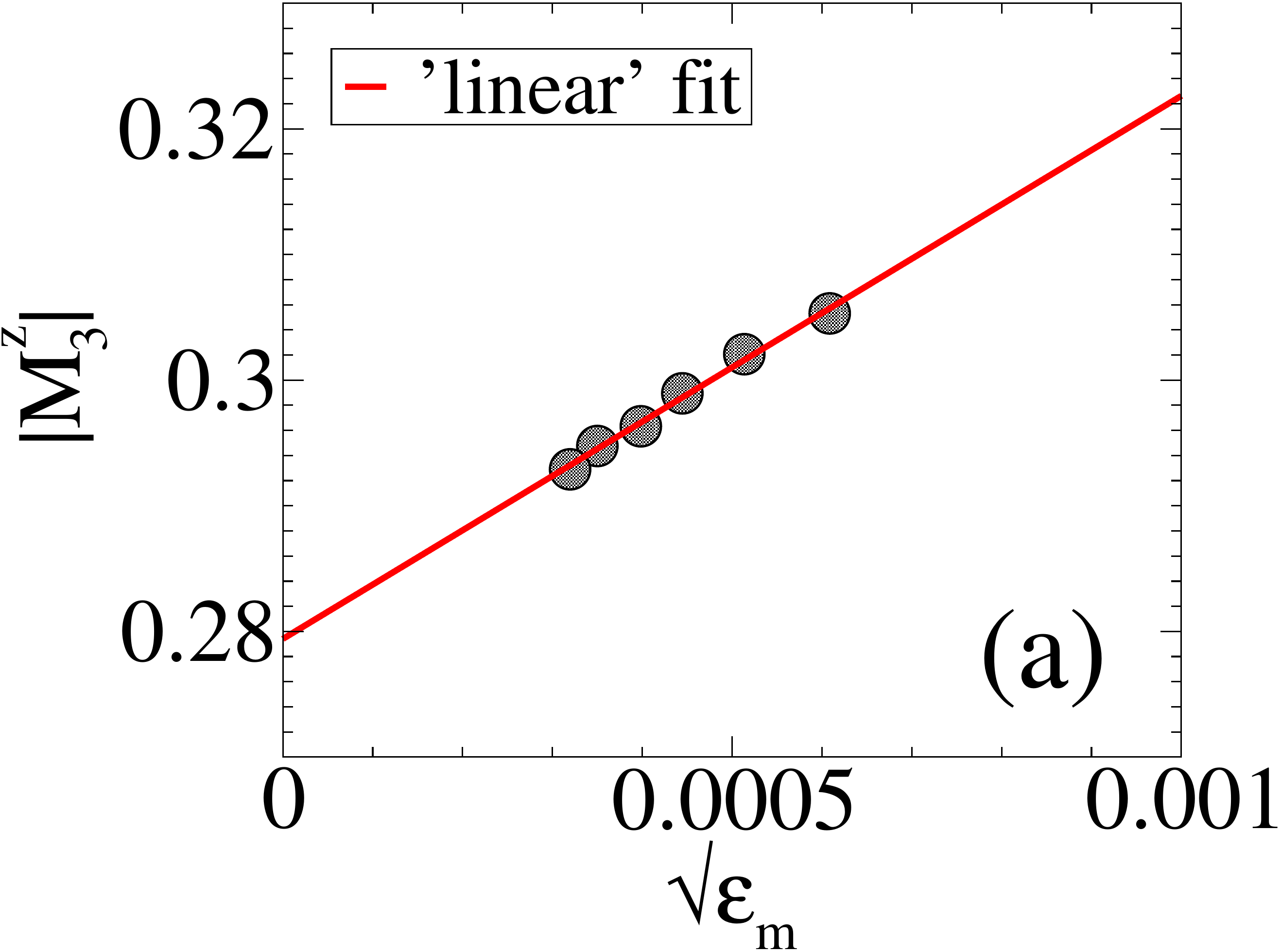}
    \includegraphics[width=0.49\columnwidth]{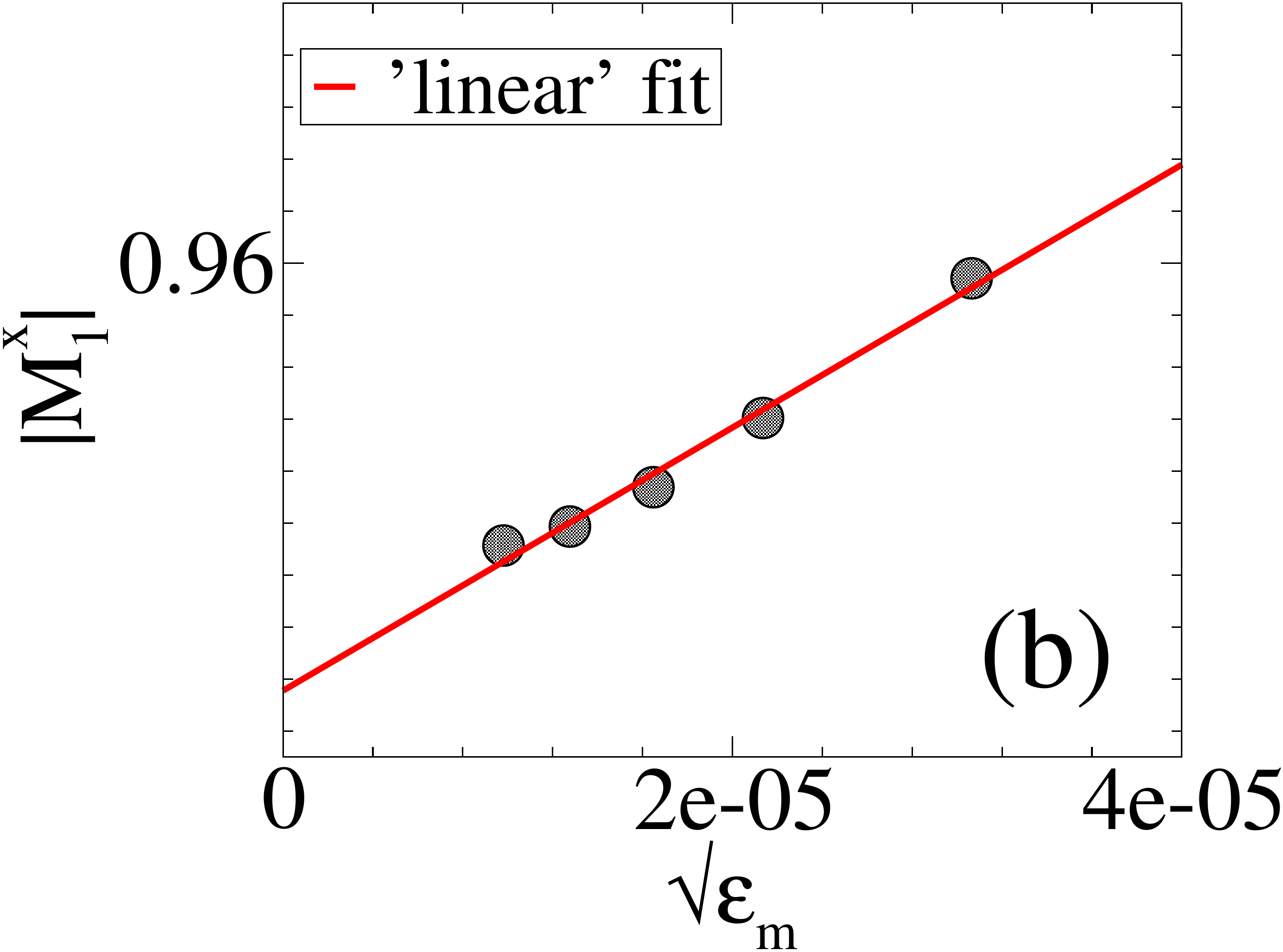} \\
    \includegraphics[width=0.49\columnwidth]{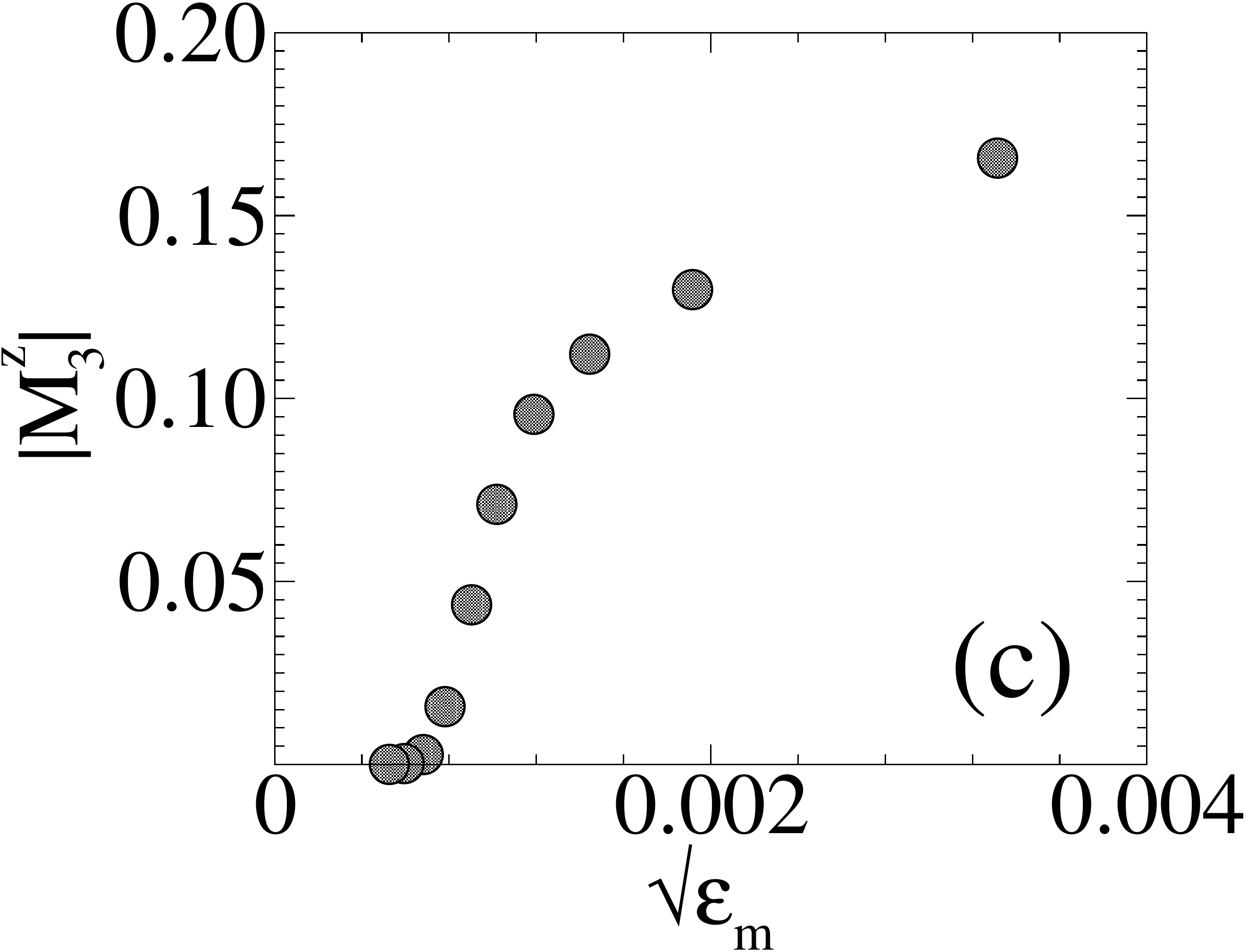}
    \includegraphics[width=0.49\columnwidth]{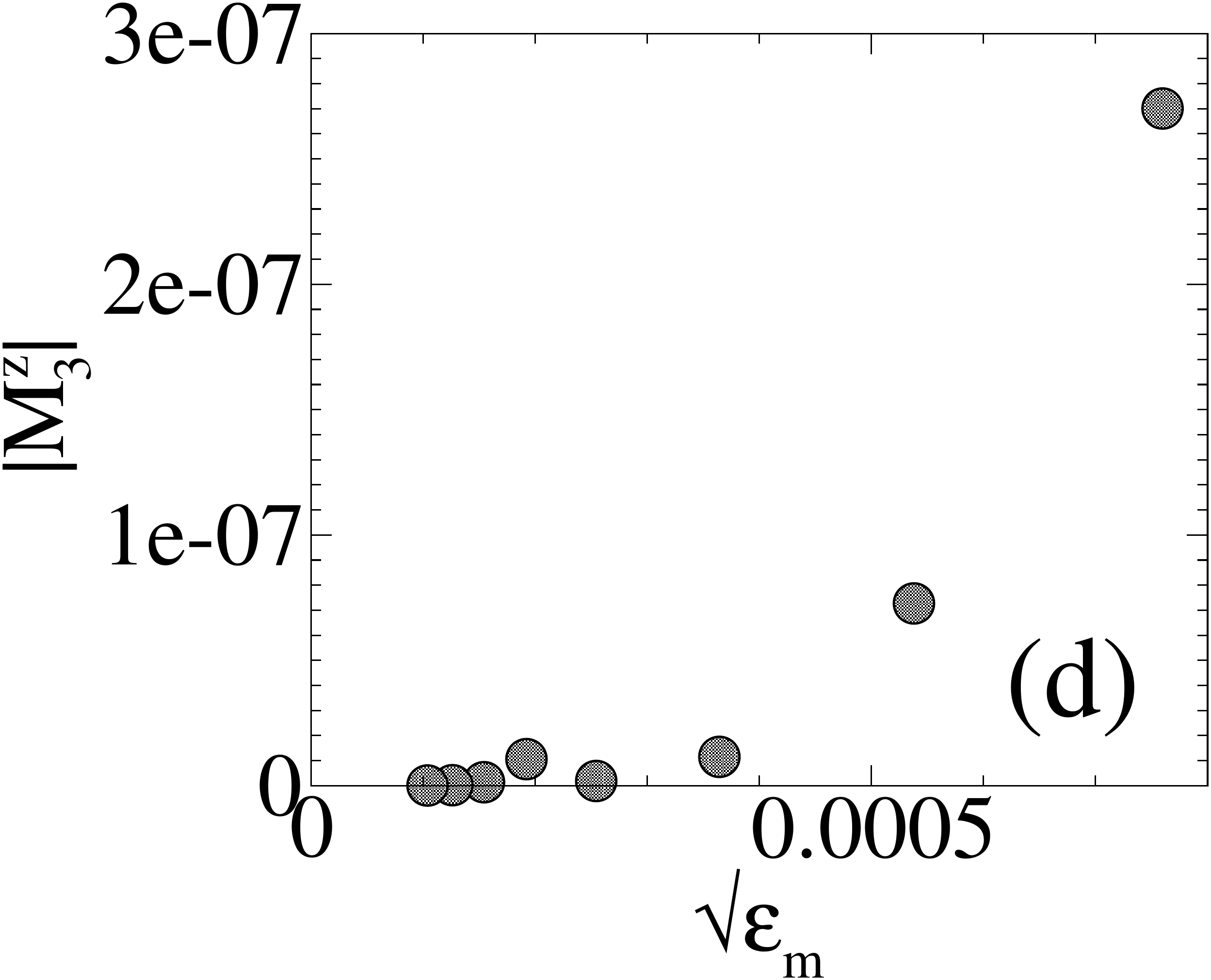}
    \caption{(Color online)
    Examples for the individual order parameter values of the ground states of
    NN-TQIM, \eref{eq:HamNN}, on YC6 structures: (a) $M_3^z$ at $\Gamma=0.5$ [deep inside $(0.5,-0.5,0)$-ordered clock phase region], 
    (b) $M_1^x$ at $\Gamma=1.75$ [deep inside $x$-polarized FM region], (c) $M_3^z$ at $\Gamma=0.75$ 
    [close to the critical point], and (d) $M_2^z$ at 
    $\Gamma=1.00$ [deep inside $x$-polarized FM region]. Evidently, for (a) and (b), rigorous linear extrapolations versus $\sqrt{\varepsilon_m}$ 
    are possible, while not for others due to extreme decay of individual magnetizations and 
    their convergence toward small values of order of the machine epsilon. 
    \label{fig:NN-MagFit-examples}}
\end{center}
\end{figure}
%%%%%%%%%%%%%%%%%%%%%%%%%%%%%%%%%%%%%%%%%%%%%%%%%%%%%%%%%%%%%%%%%%%%%%%%%%

\section{Phase diagram of the NN model}
\label{sec:PhaseDiagramNN}
% including the detailed (property) list of detected ground states.

%%% Paragraph 1: introducing the phase diagram of the NN model
Is this section, we present the iDMRG phase diagram of the NN-TQIM, with Hamiltonian given by \eref{eq:HamNN}, on 6-leg cylinders in \fref{fig:PhaseDiagramNN},
where the $S^x$-magnetization ($M_1^x$), staggered $S^z$-magnetization ($M_2^z$), and $XY$-magnetization ($M_3^z$) 
per site are plotted. For the majority of $\Gamma$-points, the extrapolations 
toward the thermodynamic limit of $m\rightarrow\infty$ are performed
linearly with $\sqrt{\varepsilon_m}$, where $\varepsilon_m$ is the average truncation error of iDMRG for a fixed-$m$
sweep, as it was suggested by White and Chernyshev~\citep{White07} for observables other than the energy (recall that
$m\rightarrow\infty$ corresponds to $\varepsilon\rightarrow0$ limit). However, since the scaling 
behaviors of observables vary unpredictably in the vicinity of a critical point or deep inside a
phase region that is paramagnetic with respect to the target order parameter, it was not possible to perform
such extrapolations everywhere. For these points, we observed that the decay of the order parameters
are too rapid and/or the individual values are too small (in order of the machine epsilon).
We replace $M_{\{(1),2,3\}}^{\{(x),z\}}(m\rightarrow\infty)$ with $M_{\{(1),2,3\}}^{\{x,z\}}(m_{\rm max})$, virtually implying 
zero uncertainty for these points. Four examples of individual magnetization values are presented in
\fref{fig:NN-MagFit-examples}, where two sub-figures correspond to large individual values of magnetizations, 
deep inside matching phase regions where a linear fit versus $\sqrt{\varepsilon_m}$ works
well, while other sub-figures correspond to $\Gamma$ close to a predicted critical point and/or where 
magnetizations are decaying too fast, and so  no analytical fit is applicable. Using this approach, we estimate that the
critical point of the NN model lies on $\Gamma_c=0.75(5)$, i.e. the first point that $M_3^z$
touches the zero axis.  This corresponds to a second-order %(continuous) 
quantum phase transition, due to observed continuous changes in the values of magnetizations which are caused by the quantum fluctuations. 
The critical point of the model on YC6 triangular-lattice structures 
is relatively close to $\Gamma_c\approx0.705$ (in our Hamiltonian notation of \eref{eq:HamNN}) 
predicted by Penson\etal~\citep{Penson79} and $\Gamma_c\approx0.825$ by Isakov and Moessner~\citep{Isakov03} (see also~\sref{sec:intro}). 
For $\Gamma<0.75(5)$, $M_3^z(m\rightarrow\infty)$-values are finite and large,
while $M_1^x(m\rightarrow\infty)$ ($M_2^z(m\rightarrow\infty)$) values are relatively (very) small, which suggests the phase is a three-sublattice
AFM clock $(0.5,-0.5,0)$-order %(Importantly, this ground state is only SR-correlated -- 
(see below for detailed properties). 
The convergence of iDMRG ground states to such a $(0.5,-0.5,0)$-order is
consistent with the proposed ground state from Ref.~\onlinecite{MoessnerSondhi01,Isakov03}. For $\Gamma\geq0.75(5)$, $M_2^z(m_{\rm max})$ 
and $M_3^z(m_{\rm max})$ are vanishing while
$M_1^x(m\rightarrow\infty)$-values are finite and large (but not equal to unity).  This behavior suggests the phase is
a partially $x$-polarized FM order, or a paramagnet considering the $z$-polarizations, as one  
expects.

%%%%%%%%%%%%%%%%%%%%%%%%%%%%%%%%%%%%%%%%%%%%%%%%%%%%%%%%%%%%%%%%%%%%%%%%%%%
\begin{figure}
  \begin{center}
    \includegraphics[width=0.49\columnwidth]{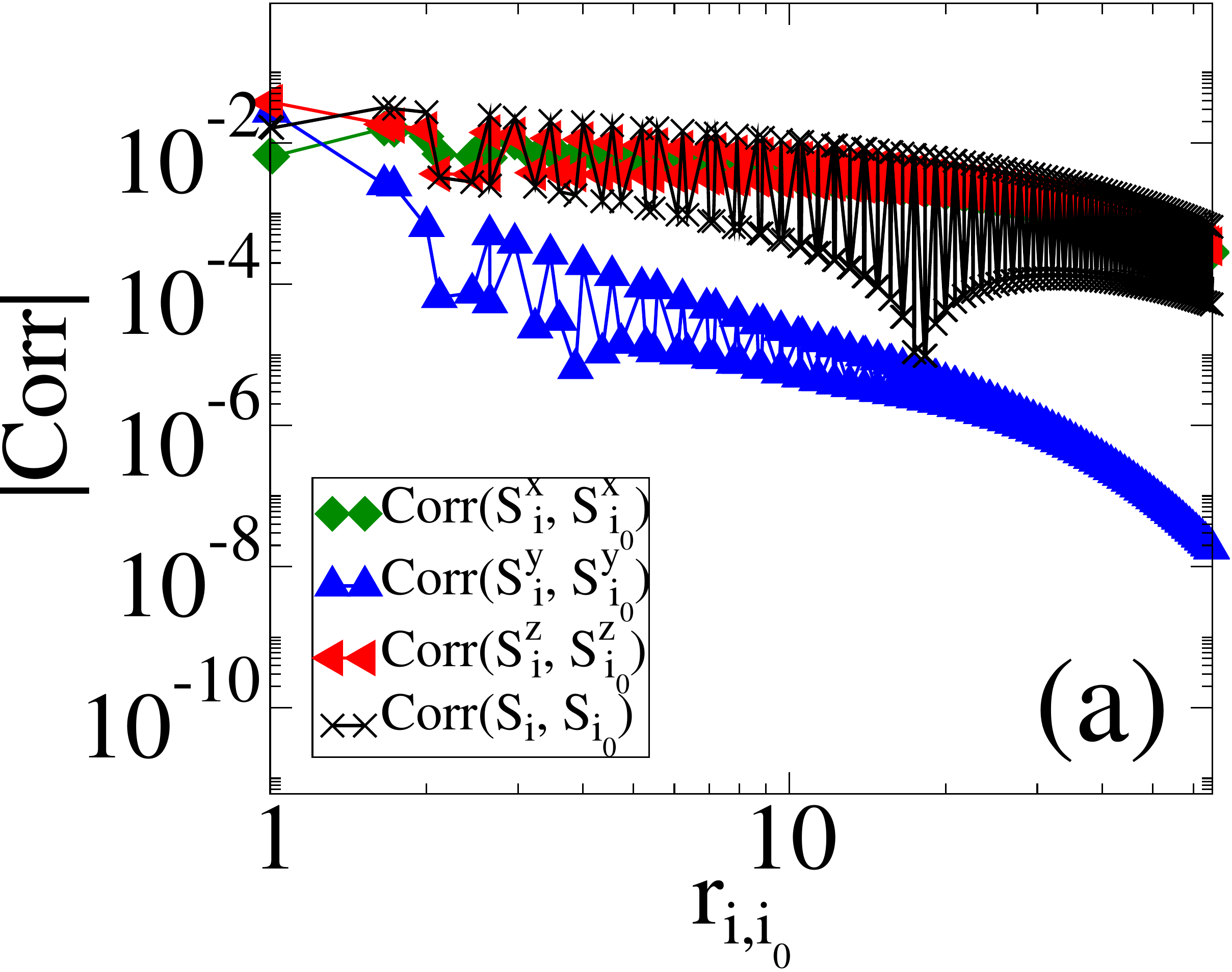}
    \includegraphics[width=0.49\columnwidth]{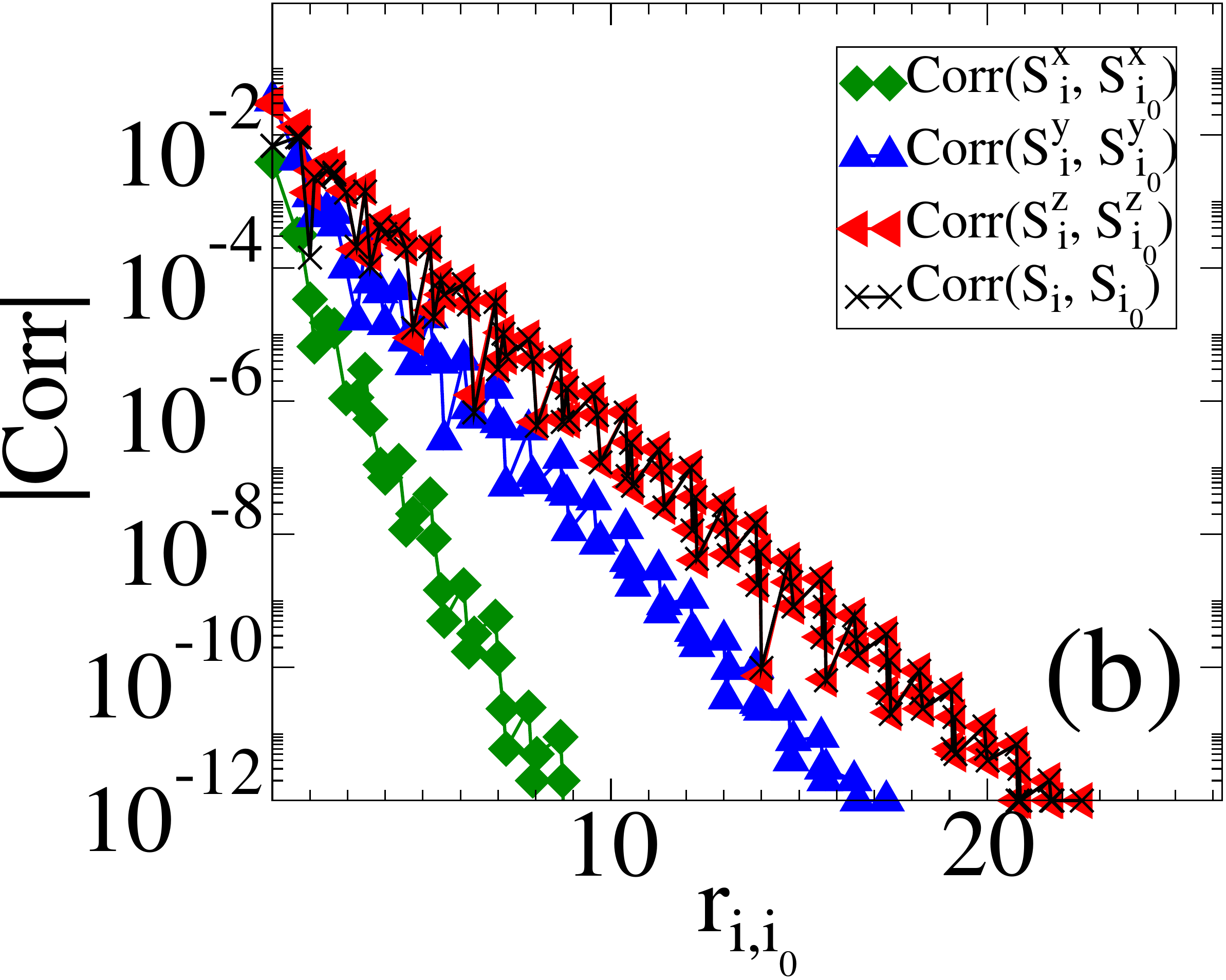}
    \caption{(Color online)
    The scaling of the connected correlation functions, $\text{Corr}(S^{\{x,y,z\}}_i,S^{\{x,y,z\}}_{i_0})$, 
    versus real-space chord distance, $r_{i,i_0}$, 
    for the iDMRG ground states of NN-TQIM, \eref{eq:HamNN}, on 
    YC6 structures at (a) $\Gamma=0.25$ [deep inside ``order from disorder''-induced $(0.5,-0.5,0)$ clock phase
    region] in \emph{full-logarithmic} scale and (b) $\Gamma=1.5$ [deep inside SR-correlated $x$-polarized FM phase region]
    in \emph{linear-log} scale.
    \label{fig:NN-CorrFunc}}
\end{center}
\end{figure}
%%%%%%%%%%%%%%%%%%%%%%%%%%%%%%%%%%%%%%%%%%%%%%%%%%%%%%%%%%%%%%%%%%%%%%%%%%

%%%%%%%%%%%%%%%%%%%%%%%%%%%%%%%%%%%%%%%%%%%%%%%%%%%%%%%%%%%%%%%%%%%%%%%%%%%
\begin{figure}
  \begin{center}
    \includegraphics[width=0.99\columnwidth]{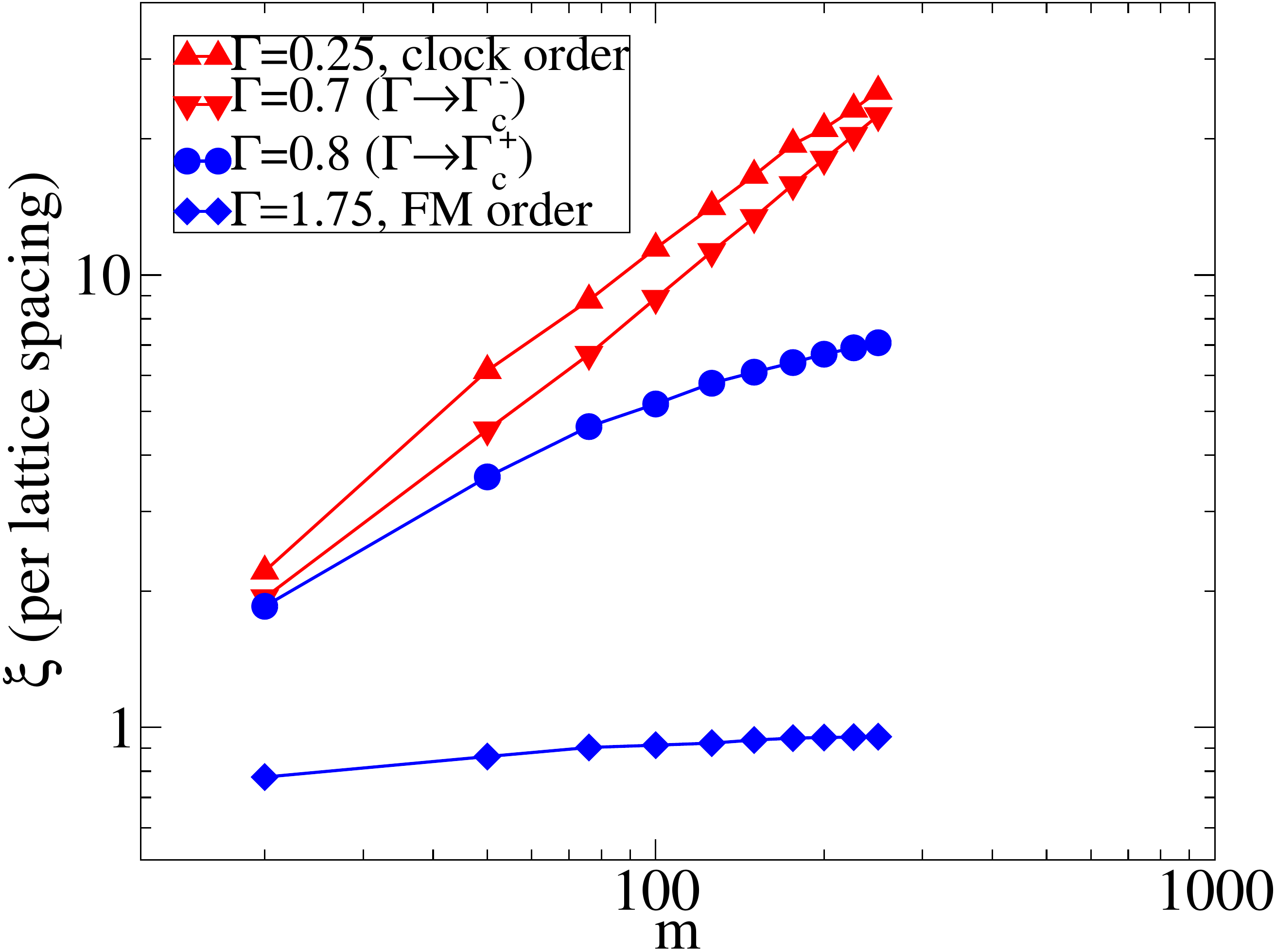}
    \caption{(Color online)
    iDMRG correlation lengths for the ground states of NN-TQIM, 
    \eref{eq:HamNN}, on YC6 structures for a selection of $\Gamma$-points. 
    \label{fig:NN-CorrLength}}
\end{center}
\end{figure}
%%%%%%%%%%%%%%%%%%%%%%%%%%%%%%%%%%%%%%%%%%%%%%%%%%%%%%%%%%%%%%%%%%%%%%%%%%

%%% Paragraph 2: detailing the properties of detected ground states for the NN model
We scrutinize the properties of detected ground states of $H_{NN}$ by 
considering some more iDMRG observables in the following list:
 %%%
\begin{enumerate}
 %%%
%%%%%%%%%%%%%%%%%%%%%%%%%%%%%%%%%%%%%%%%%%%%%%%%%%%%%%%%%%%%%%%%%%%%%%%%%%%%%%%%%%%%%%%%%%%%%%%%%%%%%%%%%%%%%
\item[1.] \textbf{$0 < \Gamma \leq 0.75(5)$, the ``order from disorder''-induced clock $(0.5,-0.5,0)$-order}:
%%%%%%%%%%%%%%%%%%%%%%%%%%%%%%%%%%%%%%%%%%%%%%%%%%%%%%%%%%%%%%%%%%%%%%%%%%%%%%%%%%%%%%%%%%%%%%%%%%%%%%%%%%%%%
  The ground state is a $Z_2$-symmetry-broken three-sublattice order and exhibits an AFM arrangement 
  of spins in a triangular plaquette according to (0.5,-0.5,0), or $(\uparrow,\downarrow,\rightarrow)$, 
  which has a zero net magnetization. The existence of this long-range spin ordering is evident from 
  finite and large $M_3^z(m\rightarrow\infty)$-values that appeared in \fref{fig:PhaseDiagramNN}; in addition, 
  we verified the $(0.5,-0.5,0)$ structure by studying the real-space visualization of correlation
  functions (not presented here). In \sref{sec:intro}(C), we have learned that the classical ground state ($\Gamma=0$)
  is a macroscopically-degenerate LR-correlated disordered phase, where quantum-to-classical mapping implies that the 
  finite-temperature states choose a LR-correlated three-sublattice order induced by classical version
  of ``order from disorder'' phenomenon. We argue that the ground state of NN-TQIM for $0 < \Gamma \leq 0.75(5)$ is the quantum
  analog of this finite-temperature phase, where one needs to replace the temperature with $\Gamma$ 
  (``order from disorder'' is now induced by quantum fluctuations) consistent with Refs.~\onlinecite{Penson79,MoessnerSondhi01,Isakov03}. Our results confirm that the clock order is LR-correlated as observed from the 
  \emph{almost} algebraic decays of two-point \emph{connected} correlation functions, 
  $\text{Corr}(S^{\mathfrak{a}}_i,S^{\mathfrak{a}}_{i_0})=\la S^{\mathfrak{a}}_i S^{\mathfrak{a}}_{i_0} \ra - \la S^{\mathfrak{a}}_i \ra \la S^{\mathfrak{a}}_{i_0} \ra$,
  $\mathfrak{a} \in \{x,y,z\}$ [and defining $\text{Corr}(\vektor{S}_i,\vektor{S}_{i_0})= \text{Corr}(S^x_i,S^x_{i_0}) + \text{Corr}(S^y_i,S^y_{i_0})
  + \text{Corr}(S^z_i,S^z_{i_0})$], shown in \fref{fig:NN-CorrFunc}(a) for $\Gamma=0.25$.
  Note that in the figure, which belongs to a $m=250$ wavefunction, 
  it may appear that for long distances the correlators start to drop
  exponentially fast; however, we suggest this is a finite-$m$ effect and for $m\rightarrow\infty$, there should exist a perfect
  power-law decay. When we decreased the number of states, 
  the exponential-drop tail did appear, and always at shorter distances. Moreover, a power-law growth of correlation 
  lengths versus $m$ is observed for this order as shown in \fref{fig:NN-CorrLength} for $\Gamma=0.25$. Although
  this ground state resembles 1D critical phases by possessing 
  an algebraic increase of the correlation lengths up to $\xi(m_{max}) \sim O(10)$ per Hamiltonian unit-cell size, we predict its 
  stabilization here is an inherently 2D phenomenon.  
  %This suggests the existence of LR entanglements, however, the correlation lengths may still saturate for larger $m$. 
  %The symmetry breaking structure and the results from Ref.~\onlinecite{MoessnerSondhi01} suggest 
  %that this phase has a finite spin gap.
 %%%
%%%%%%%%%%%%%%%%%%%%%%%%%%%%%%%%%%%%%%%%%%%%%%%%%%%%%%%%%%%%%%%%%%%%%%%%%%%%%%%%%%%%%%%
\item[2.] \textbf{$\Gamma \geq 0.75(5)$, the SR-correlated $x$-polarized FM order}:
%%%%%%%%%%%%%%%%%%%%%%%%%%%%%%%%%%%%%%%%%%%%%%%%%%%%%%%%%%%%%%%%%%%%%%%%%%%%%%%%%%%%%%% 
  The observed ground state exhibit partially polarized spins that are ferromagnetically aligning in spin's $x$-direction while possessing 
  vanishing magnetization (i.e.,~exhibiting paramagnetism) in other directions. We verified the FM structure by observing finite and large
  values of $M_1^x(m\rightarrow\infty)$ (non-zero net magnetization) and 
  vanishing values of $M_2^z(m_{max})$ and $M_3^z(m_{max})$ as shown in~\fref{fig:PhaseDiagramNN}.  This was also supported through
  visualizations of real-space correlation functions (not presented here). The FM order is SR-correlated 
  due to exponentially-decaying connected correlators, as shown in \fref{fig:NN-CorrFunc}(b)
  for $\Gamma=1.5$, and therefore gapped and SR-entangled due to small and saturating
  correlation lengths (when plotting versus $m$), as shown in \fref{fig:NN-CorrLength} for, e.g., $\Gamma=1.75$. 
  %The FM order posses spin-flip gapped elementary excitations, as a version of Hastings theorem consistently 
  %predict finite bulk gap for observing saturating correlation lengths. 
 %%%
\end{enumerate}

\section{Phase diagram of the LR model}
\label{sec:PhaseDiagramLR}
% including the detailed (property) list of detected ground states.

%%%%%%%%%%%%%%%%%%%%%%%%%%%%%%%%%%%%%%%%%%%%%%%%%%%%%%%%%%%%%%%%%%%%%%%%%%%
\begin{figure}
  \begin{center}
    \includegraphics[width=0.99\columnwidth]{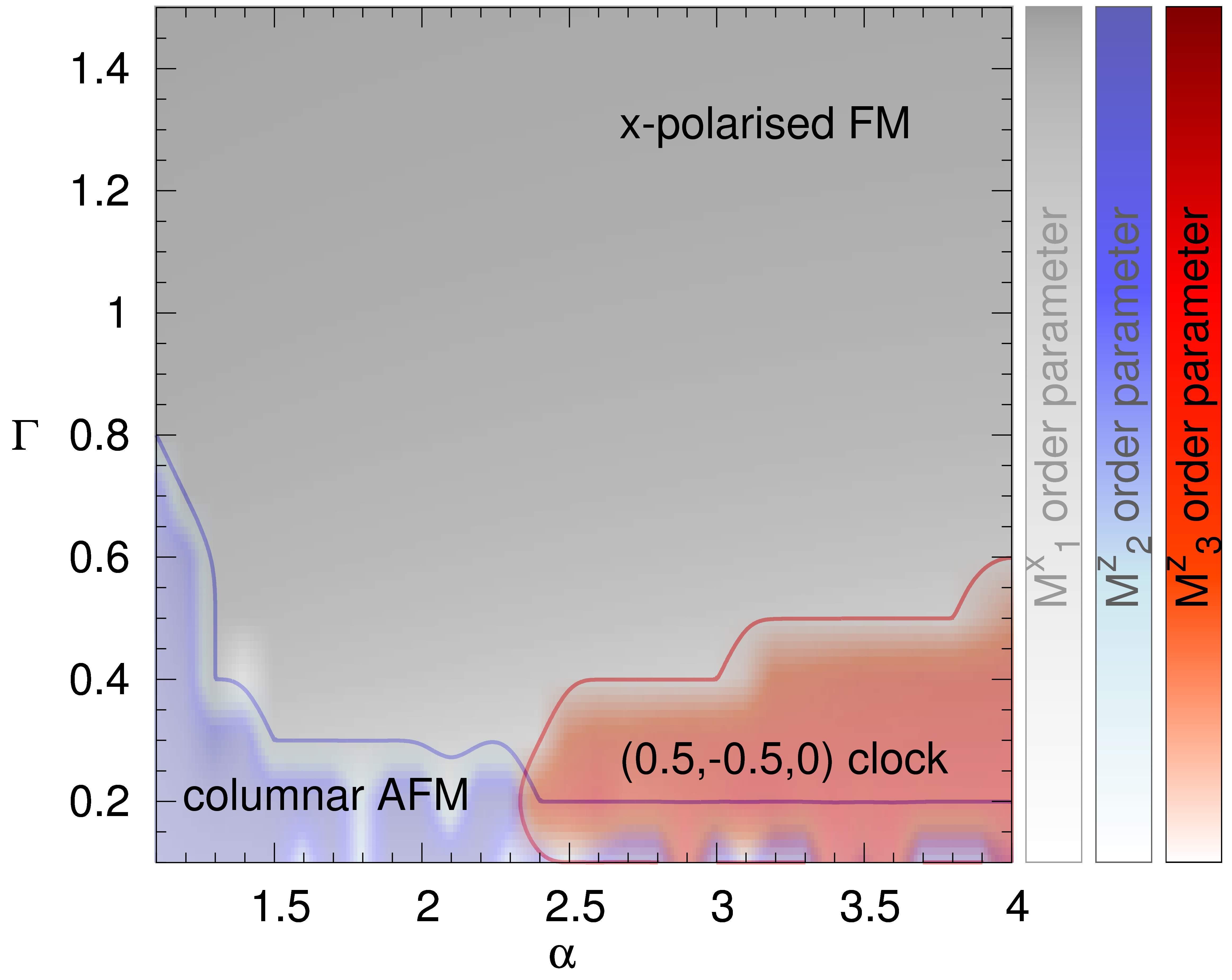}
    \caption{(Color online)
    iDMRG phase diagram of LR-TQIM, \eref{eq:HamLR}, on a YC6 structure. This shows the combined 
    color map of measured magnetizations per site, namely $M_1^x$, $M_2^z$, and $M_3^z$ 
    [cf.~Eqs.~\ref{eq:M1x}, \ref{eq:M2z}, and \ref{eq:M3z} respectively -- we set $N_{xy}=\frac{1}{\sqrt{12}}$ for $M_3^z$ as appeared in \eref{eq:Oxy}].
    The color intensity of all palettes vary in the range of $[0,1]$, as expected for normalized order parameters.
    The thick blue (red) line is the zero-value [while taking into account the maximal uncertainty in measurements of the 
    magnetizations] contour line specifying the phase boundary between the LR-correlated columnar
    order (clock $(0.5,-0.5,0)$-order) and other ground states. Here, for the majority of $(\alpha,\Gamma)$-points, we
    insert the thermodynamic-limit magnetizations, $M_1^x(m\rightarrow\infty)$, $M_2^z(m\rightarrow\infty)$, 
    and $M_3^z(m\rightarrow\infty)$, which are extrapolated using a linear fit versus $\sqrt{\varepsilon_m}$ (see below for some examples 
    on individual fits). However, where no analytical
    fit is possible (due to extreme decay and/or smallness of observables), we 
    instead insert $M_1^x(m_{\max})$, $M_2^z(m_{\max})$, and $M_3^z(m_{\max})$ as needed.
    \label{fig:PhaseDiagramLR}}
\end{center}
\end{figure}
%%%%%%%%%%%%%%%%%%%%%%%%%%%%%%%%%%%%%%%%%%%%%%%%%%%%%%%%%%%%%%%%%%%%%%%%%%

%%% Paragraph 1: introducing the phase diagram of the LR model
The fully-quantitative iDMRG phase diagram of LR-TQIM, with Hamiltonian given by \eref{eq:HamLR}, is presented in \fref{fig:PhaseDiagramLR}. 
This figure displays the three normalized
order parameters of interest, i.e.~$M_1^x$, $M_2^z$, and $M_3^z$ [cf.~eqs.~\ref{eq:M1x}, \ref{eq:M2z}, and \ref{eq:M3z} respectively], 
corresponding to the stabilization of three observed 
ground states: LR-correlated $x$-polarized FM, LR-correlated columnar AFM, and LR-correlated
clock $(0.5,-0.5,0)$ order, respectively.

%%%%%%%%%%%%%%%%%%%%%%%%%%%%%%%%%%%%%%%%%%%%%%%%%%%%%%%%%%%%%%%%%%%%%%%%%%%
\begin{figure}
  \begin{center}
    \includegraphics[width=0.49\columnwidth]{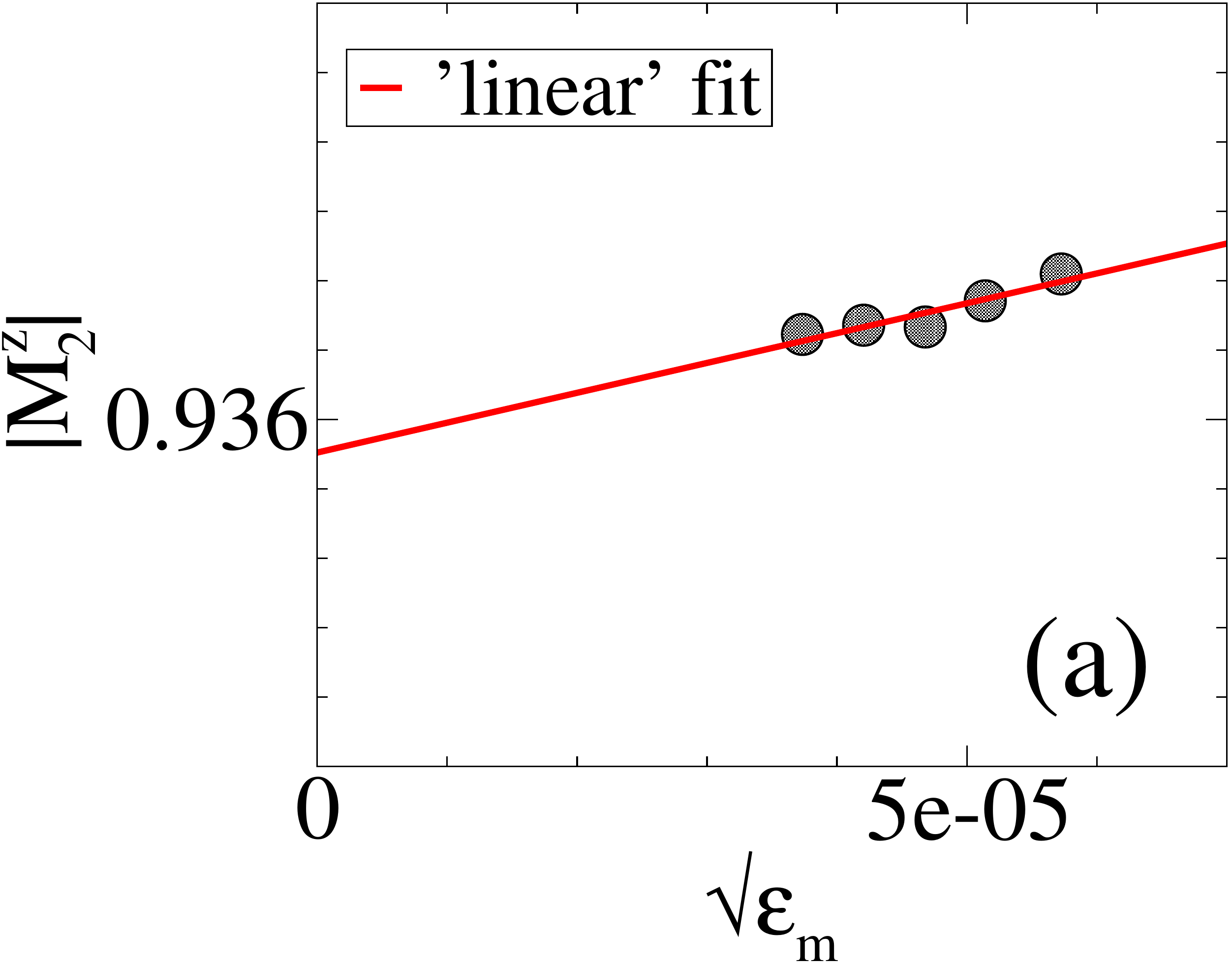}
    \includegraphics[width=0.49\columnwidth]{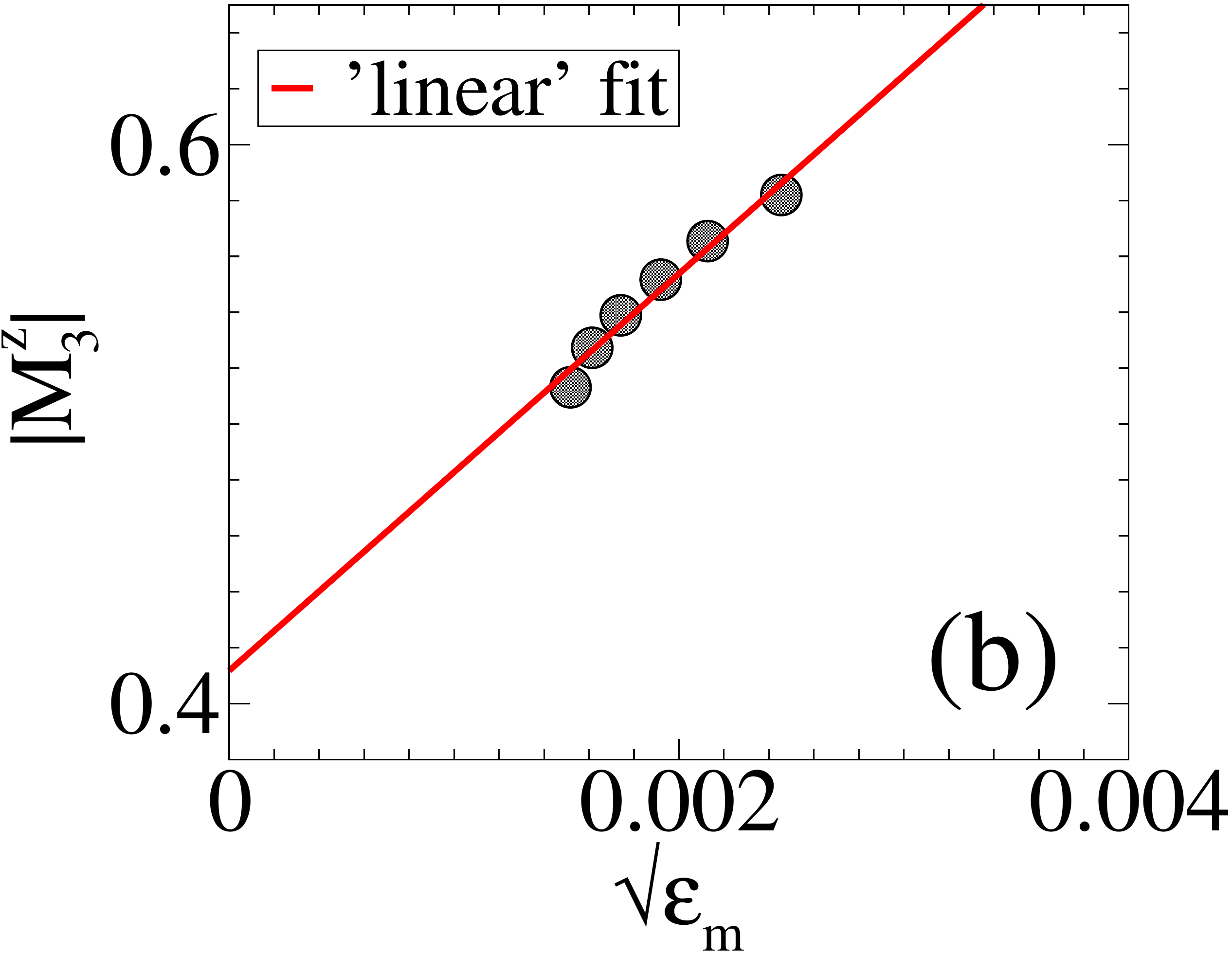} \\
    \includegraphics[width=0.49\columnwidth]{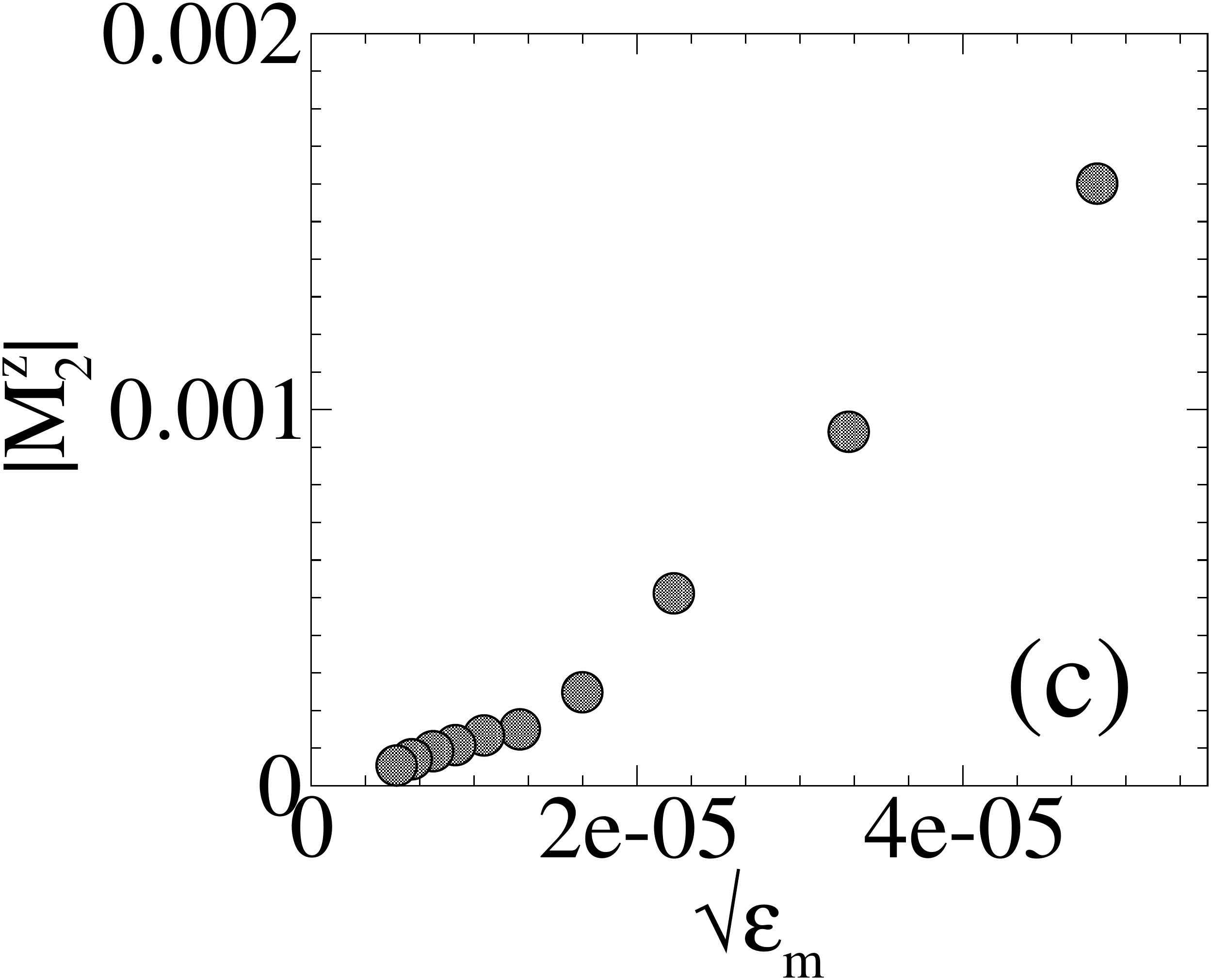}
    \includegraphics[width=0.49\columnwidth]{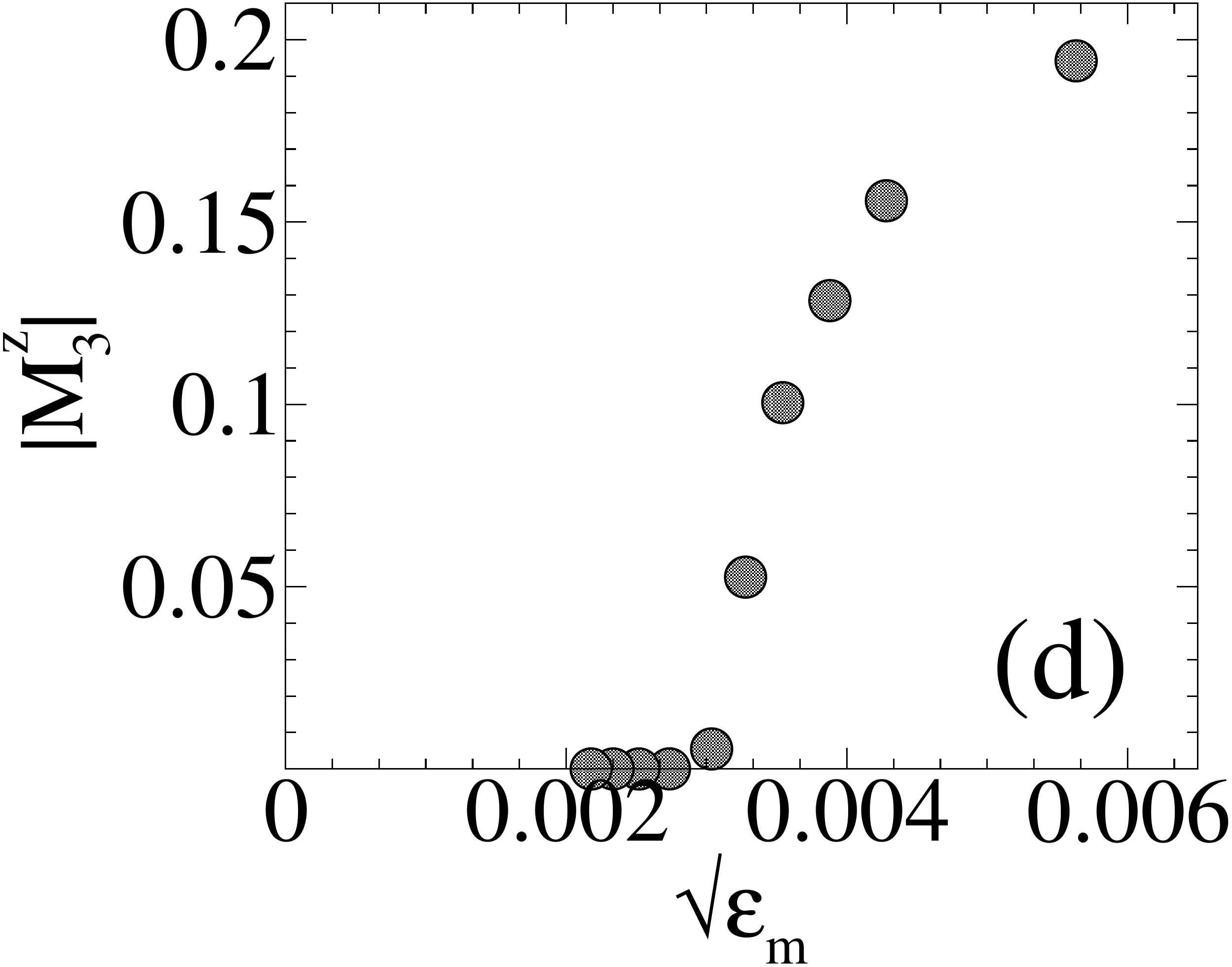}
    \caption{(Color online)
    Examples for the individual order parameter values for the ground states of
    LR-TQIM, \eref{eq:HamLR}, on YC6 structures: (a) $M_2^z$ at $(\alpha,\Gamma)=(1.6,0.2)$ [deep inside columnar AFM phase region], 
    (b) $M_3^z$ at at $(\alpha,\Gamma)=(2.6,0.3)$ [deep inside $(0.5,-0.5,0)$ clock phase region], (c) $M_2^z$ at 
    $(\alpha,\Gamma)=(1.5,0.3)$ [close to a critical point], and (d) $M_3^z$ at 
    $(\alpha,\Gamma)=(2.2,0.3)$ [close to a critical point]. Evidently, for (a) and (b), rigorous linear 
    extrapolations versus $\sqrt{\varepsilon_m}$ are possible, while not for others due to extreme decay of individual magnetizations and 
    their convergence toward small values of order of the machine epsilon.  
    \label{fig:LR-MagFit-examples}}
\end{center}
\end{figure}
%%%%%%%%%%%%%%%%%%%%%%%%%%%%%%%%%%%%%%%%%%%%%%%%%%%%%%%%%%%%%%%%%%%%%%%%%%

%%% Paragraph 1-2:
Analogous to the NN phase diagram shown in \fref{fig:PhaseDiagramNN}, for the majority of control parameters in \fref{fig:PhaseDiagramLR}
it was possible to perform the linear extrapolation of the
magnetizations versus $\sqrt{\varepsilon_m}$ toward the thermodynamic limit of $m\rightarrow\infty$; however, as before, 
typically close to critical lines or deep inside a paramagnetic phase (with respect to the targeted order parameter),
there exist some points where no analytical fit is possible due to extreme decays of order parameters 
and/or exhibiting magnitudes as small as the machine epsilon.  In such cases,
we replace $M_{\{(1),2,3\}}^{\{(x),z\}}(m\rightarrow\infty)$ with $M_{\{(1),2,3\}}^{\{(x),z\}}(m_{max})$ 
implying strictly zero uncertainties. For some examples, \fref{fig:LR-MagFit-examples} illustrates
four plots of individual magnetizations with some different scaling behaviors.

%%% Paragraph 1-3:
In the phase diagram, \fref{fig:PhaseDiagramLR}, the two contour lines 
provide our estimations for the phase boundaries. All are predicted to be second order phase transitions. Briefly, strong $x$-polarized FM order exists for large
$\Gamma$ regardless of the values of $\alpha$; columnar order exists for small $\alpha$ and $\Gamma$; and
$(0.5,-0.5,0)$-type clock order exists for large $\alpha$ and small $\Gamma$. In addition, the coexistence 
of a weak columnar and a weak $(0.5,-0.5,0)$ order observed for $\alpha \geq 2.40(5)$ and $\Gamma \leq 0.20(5)$. 

The recent mean-field/QMC study~\citep{Humeniuk16} of the model similarly found a 
three-region semi-quantitative phase diagram having phase transition lines relatively close to our predictions.  
We note two points of distinction in our conclusions.
%while also exhibiting an $x$-polarized FM order for large $\Gamma$. 
First, Humeniuk~\citep{Humeniuk16} does not discuss the nature 
of the ground state for small $\alpha$ and $\Gamma$ (it is labeled in Ref.~\onlinecite{Humeniuk16} as a 
`classical phase').  Second, for large $\alpha$ and small $\Gamma$, they find a different type of clock phase, 
specifically, the $(0.5,-0.25,-0.25)$ ordering (the so-called $120^\circ$-ordered arrangement on a triangular plaquette),
in contrast to our results. Importantly, 
%Humeniuk's argued type of the clock-ordered ground state \emph{cannot} be trusted, since  
the stabilization of the clock $(0.5,-0.5,0)$-order on the NN model is confirmed by
\sref{sec:PhaseDiagramNN} and Ref.~\onlinecite{MoessnerSondhi01} results for small $\Gamma$; the LR model \emph{must} reproduce
the $H_{NN}$ ground state for $\alpha\rightarrow\infty$.
% Frstly, we deem necessary to clarify that for small-$\Gamma$ ground states 
%of $H_{LR}$, the `classical' label could be misleading when considering the true 2D limit: 
% such phases are \emph{not} technically the reminiscent of any known classical phases/phenomena %(to be considered semi-classical) 
% in the sense that as soon as turning on the field, $\Gamma\neq0$, 
% we observed them to possess LR correlations and LR entanglement (i.e.~acquiring inherently non-product forms).
% Moreover, for large $\alpha$ (where the clock order lives, but also, a week columnar phase can coexist according to \fref{fig:PhaseDiagramLR}),
Moreover, for large $\alpha$ and vanishing $\Gamma$, although, we already know that in the thermodynamic limit there exists a
macroscopically-degenerate finite-entropy classical ground state and any \emph{finite} $\Gamma$ would
allow quantum fluctuations to choose a distinct phase (as for $(0.5,-0.5,0)$-order of $H_{NN}$ or large-$\alpha$ order of $H_{LR}$) 
through ``order from disorder'' [cf.~\sref{sec:intro}]. But, on 
the restricted geometry of the YC structure, it appears the classical ground states are distinct: 
%if one ignores the LR correlation/entanglement patterns, columnar order is a semi-classical ground state in one sense: 
employing full diagonalization calculations for
classical $H_{LR}$ [$\Gamma=0$] on small $L_x=3,4$-length YC6 systems, for all $\alpha$, we 
detect a product-state columnar order as the lowest energy state.

%%%%%%%%%%%%%%%%%%%%%%%%%%%%%%%%%%%%%%%%%%%%%%%%%%%%%%%%%%%%%%%%%%%%%%%%%%%
\begin{figure}
 \begin{center}
    \includegraphics[width=0.99\columnwidth]{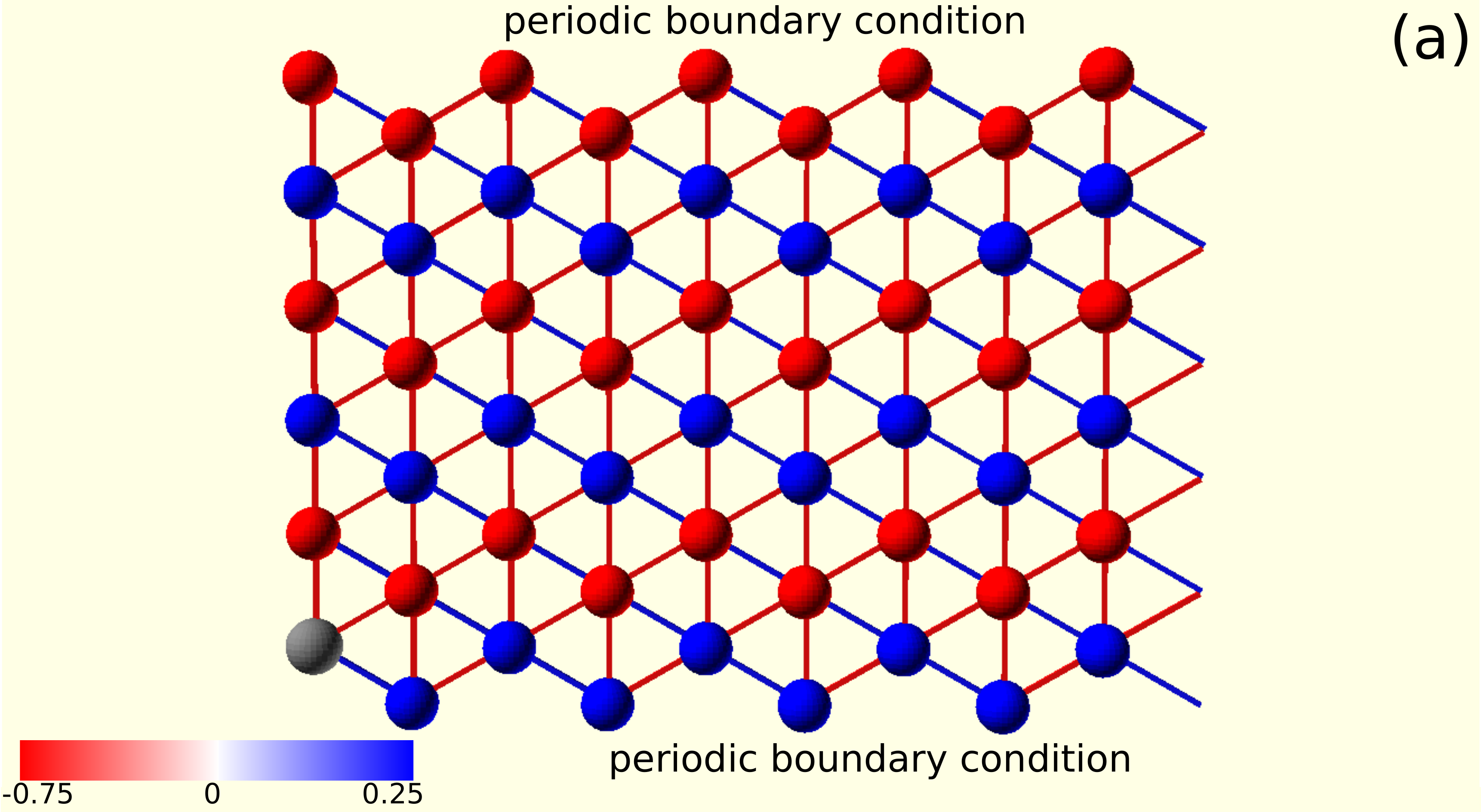}
    \includegraphics[width=0.99\columnwidth]{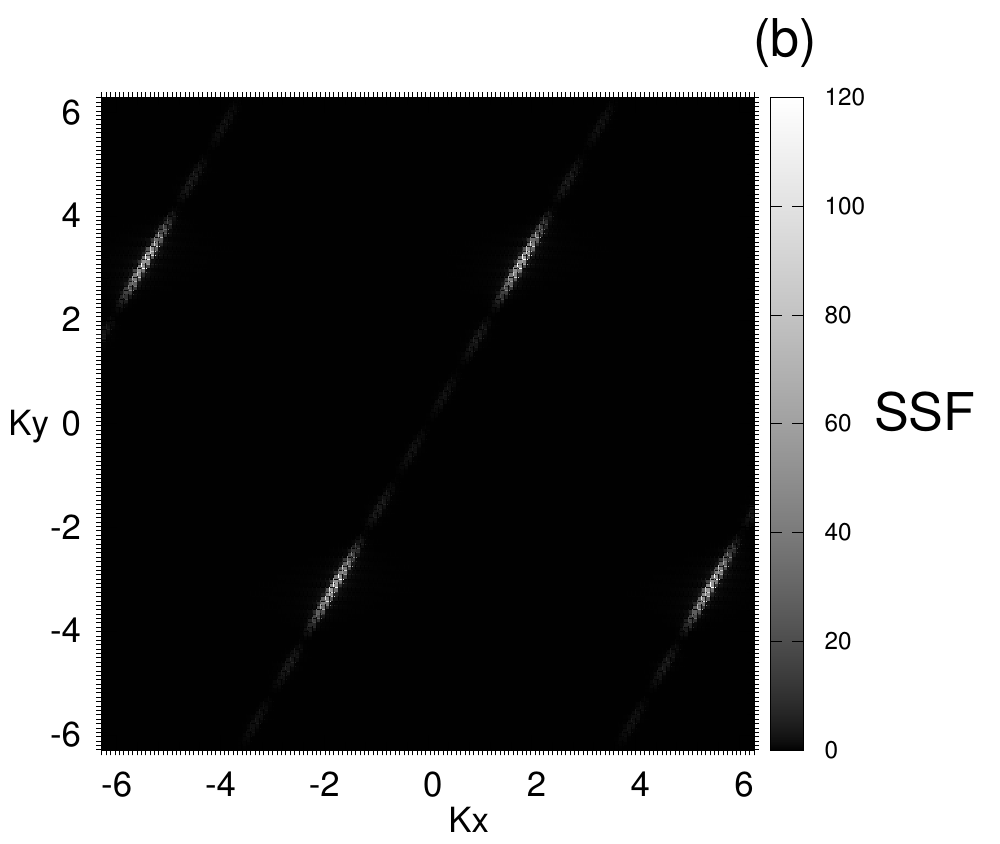}
    \caption{(Color online)
    Lattice visualizations for the iDMRG ground state
    of the LR-TQIM, \eref{eq:HamLR}, on an YC6 structure at $(\alpha,\Gamma)=(1.2,0.3)$ [LR-correlated columnar order]. 
    (a) $S^z$-$S^z$ correlation functions for up to 9 legs of the infinite cylinder. The 
    size and the color of the spheres indicate the (long-range) spin-spin correlations 
    in respect to the principal (gray) site, and the thickness and the color of the 
    bonds indicate the strength of the NN correlations. (b) SSF, where we present the
    Bragg-type peaks within the first and second Brillouin zones of the inverse lattice.
    \label{fig:VisualizationsColumnar}}
 \end{center}
\end{figure}
%%%%%%%%%%%%%%%%%%%%%%%%%%%%%%%%%%%%%%%%%%%%%%%%%%%%%%%%%%%%%%%%%%%%%%%%%%%

%%%%%%%%%%%%%%%%%%%%%%%%%%%%%%%%%%%%%%%%%%%%%%%%%%%%%%%%%%%%%%%%%%%%%%%%%%%
\begin{figure*}
  \begin{center}
    \includegraphics[width=0.32\linewidth]{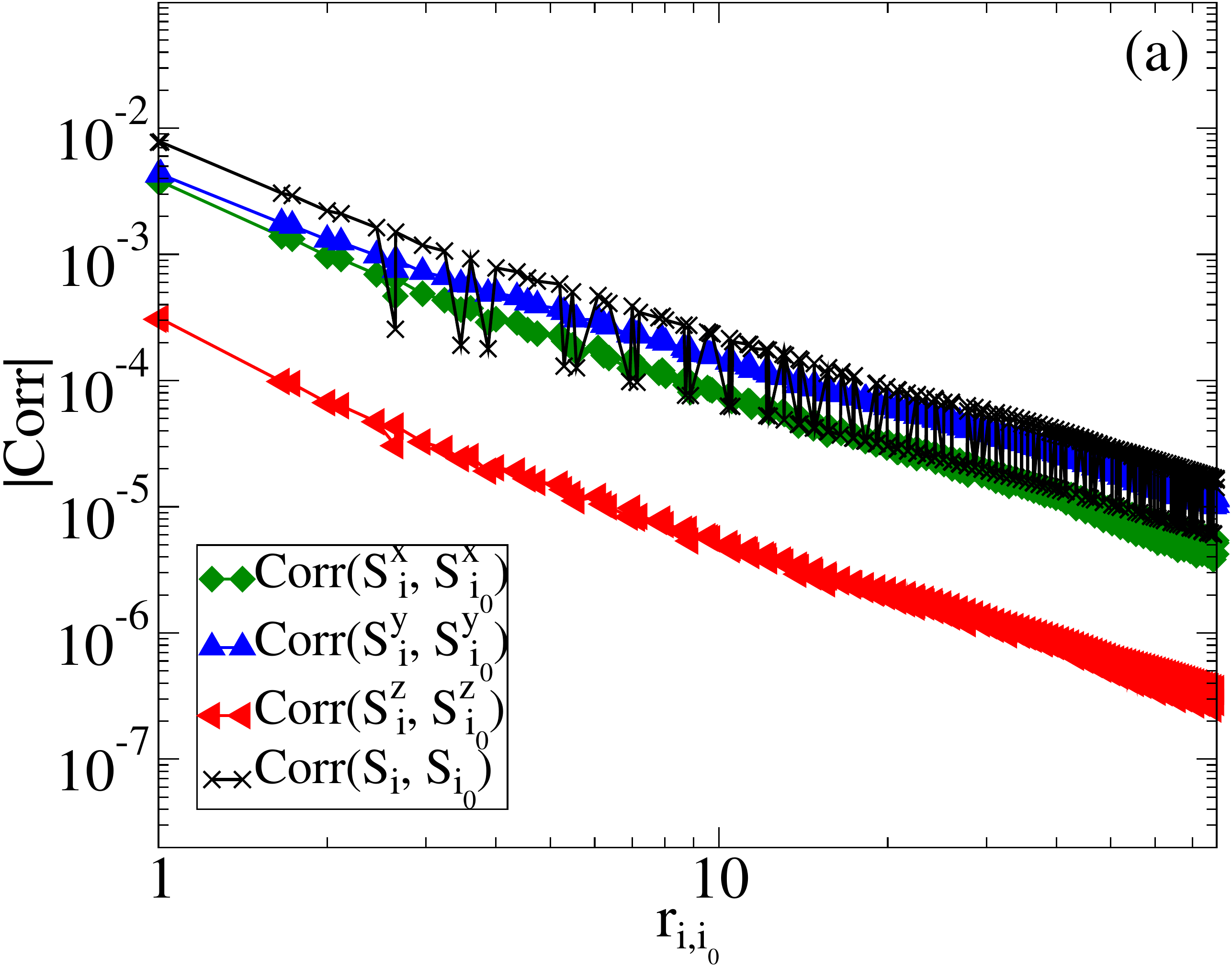}
    \includegraphics[width=0.32\linewidth]{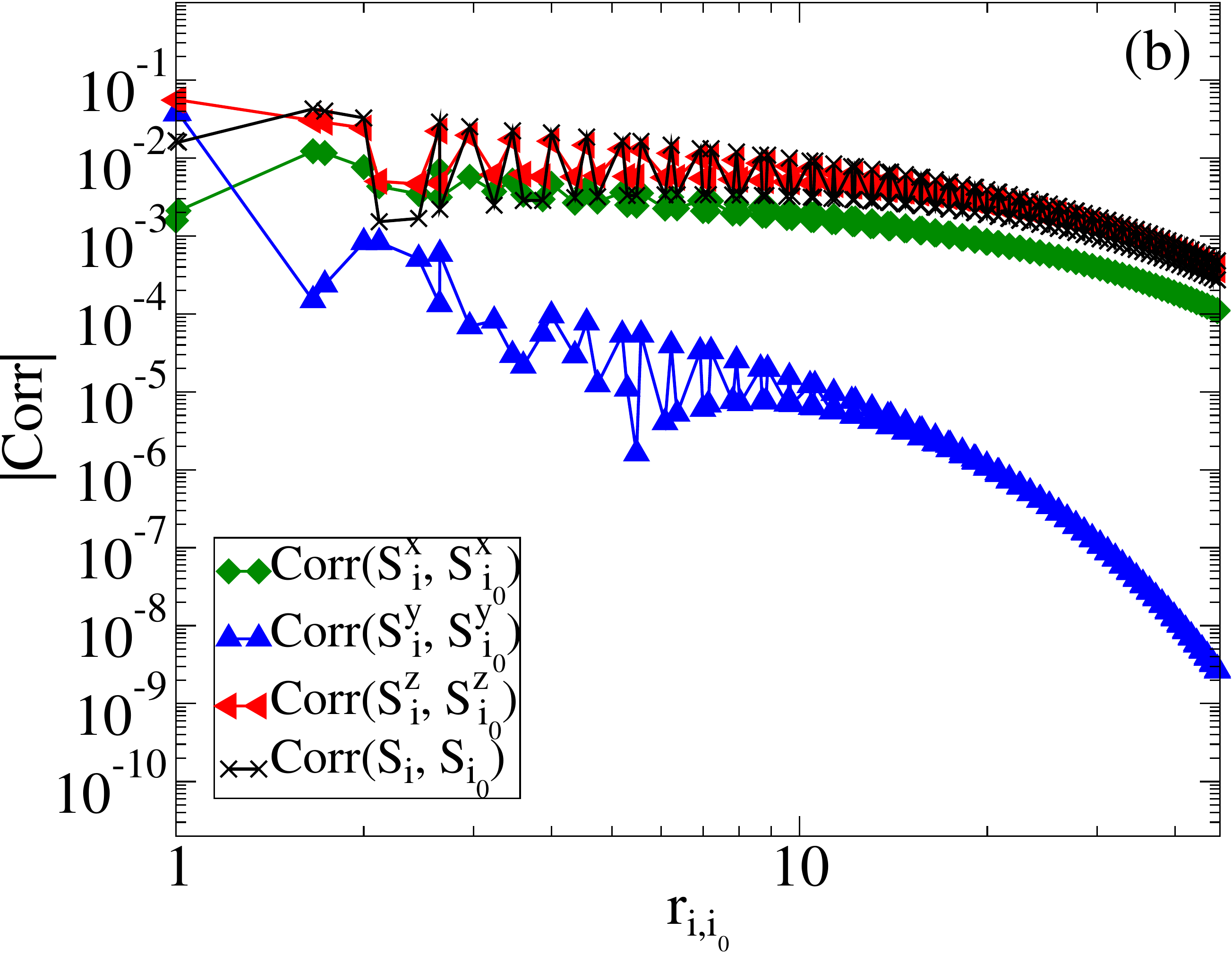}
    \includegraphics[width=0.32\linewidth]{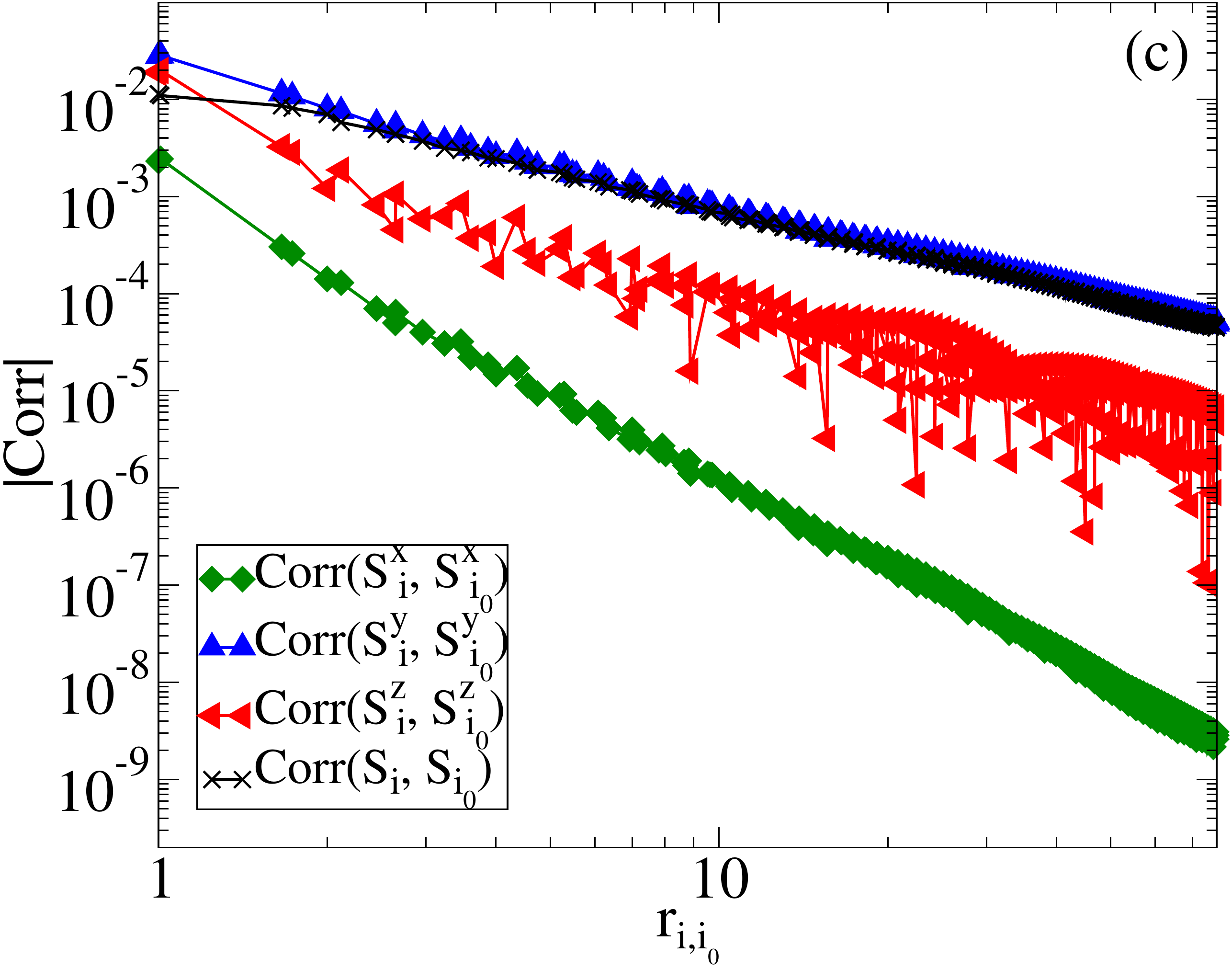}
    \caption{(Color online)
    The scaling of the connected correlation functions, $\text{Corr}(S^{\{x,y,z\}}_i,S^{\{x,y,z\}}_{i_0})$, 
    versus real-space chord distance, $r_{i,i_0}$, 
    for the iDMRG ground states of LR-TQIM, \eref{eq:HamLR}, on 
    infinite YC6 structures at (a) $(\alpha,\Gamma)=(1.2,0.3)$ [deep inside LR-correlated columnar phase
    region], (b) $(\alpha,\Gamma)=(4.0,0.3)$ [deep inside LR-correlated $(0.5,-0.5,0)$ clock phase region], and 
    (c)$(\alpha,\Gamma)=(1.2,1.5)$ [deep inside LR-correlated $x$-polarized FM phase region]. Plots are 
    in \emph{full-logarithmic} scales.
    \label{fig:LR-CorrFunc}}
\end{center}
\end{figure*}
%%%%%%%%%%%%%%%%%%%%%%%%%%%%%%%%%%%%%%%%%%%%%%%%%%%%%%%%%%%%%%%%%%%%%%%%%%

%%%%%%%%%%%%%%%%%%%%%%%%%%%%%%%%%%%%%%%%%%%%%%%%%%%%%%%%%%%%%%%%%%%%%%%%%%%
\begin{figure}
  \begin{center}
    \includegraphics[width=0.99\columnwidth]{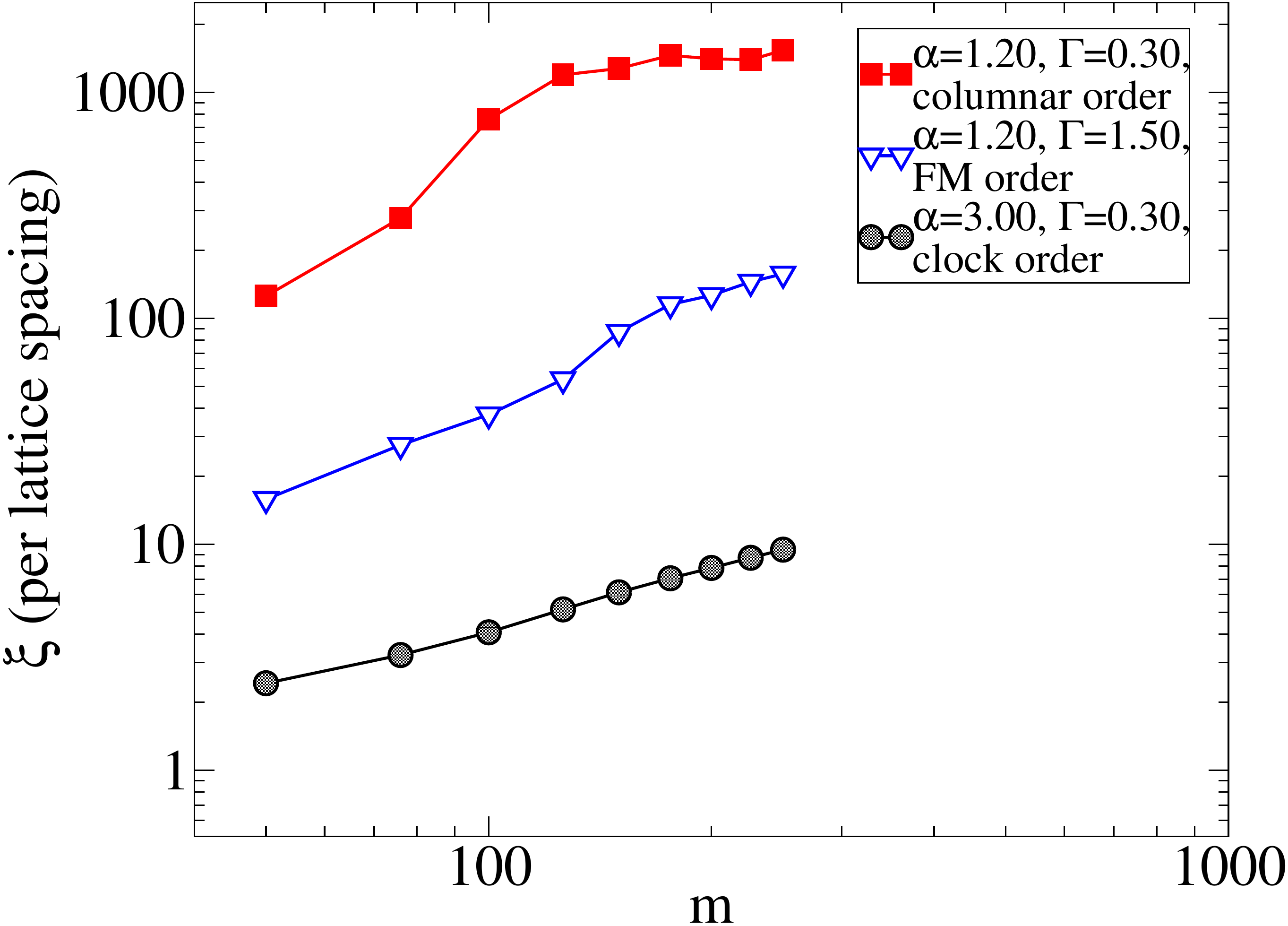}
    \caption{(Color online)
    iDMRG correlation lengths for the ground states of LR-TQIM, 
    \eref{eq:HamLR}, on infinite YC6 structures for a selection of $(\alpha,\Gamma)$-points. 
    \label{fig:LR-CorrLength}}
\end{center}
\end{figure}
%%%%%%%%%%%%%%%%%%%%%%%%%%%%%%%%%%%%%%%%%%%%%%%%%%%%%%%%%%%%%%%%%%%%%%%%%%

%%%%%%%%%%%%%%%%%%%%%%%%%%%%%%%%%%%%%%%%%%%%%%%%%%%%%%%%%%%%%%%%%%%%%%%%%%%
\begin{figure}
 \begin{center}
    \includegraphics[width=0.99\columnwidth]{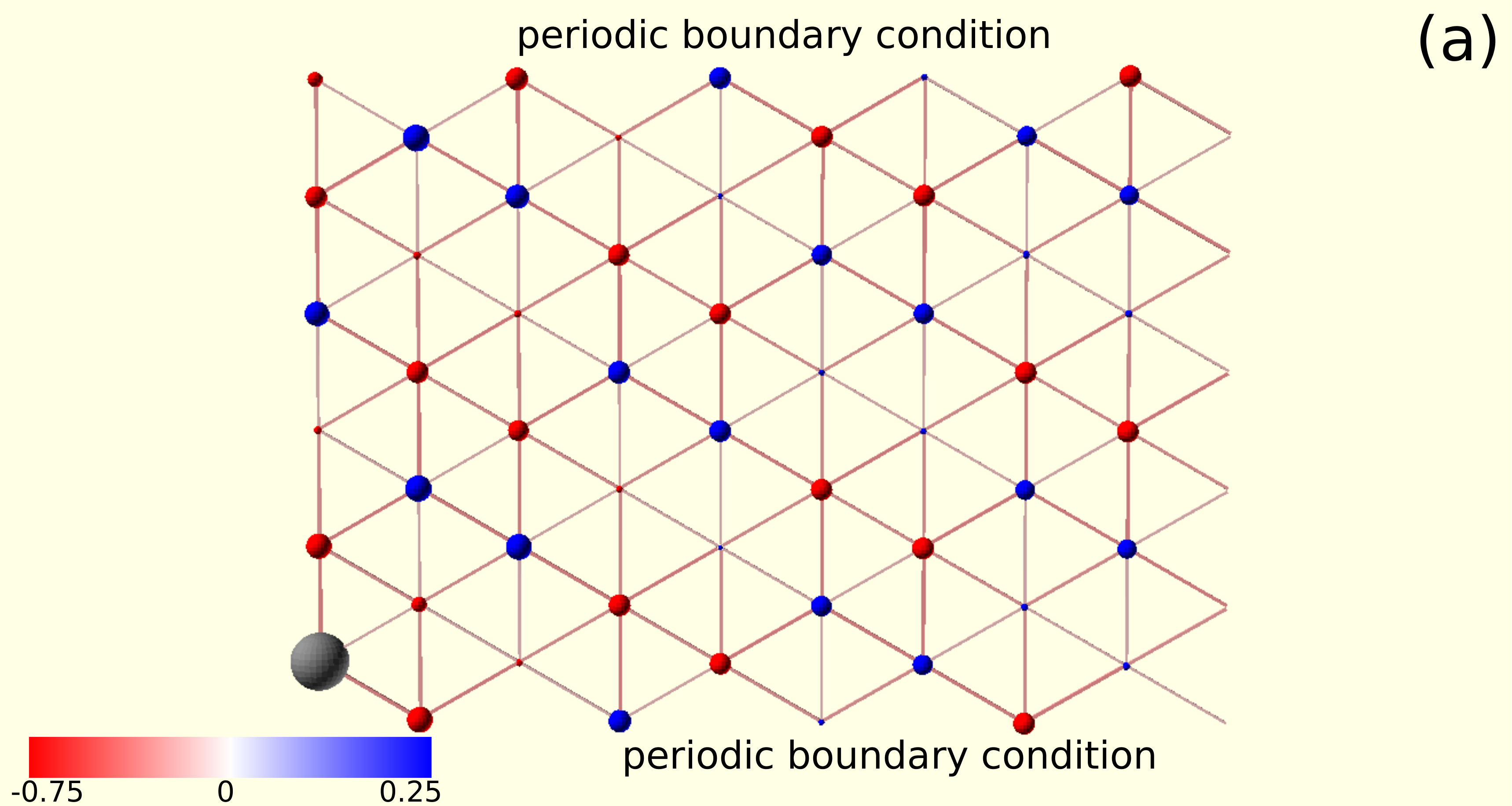}
    \includegraphics[width=0.99\columnwidth]{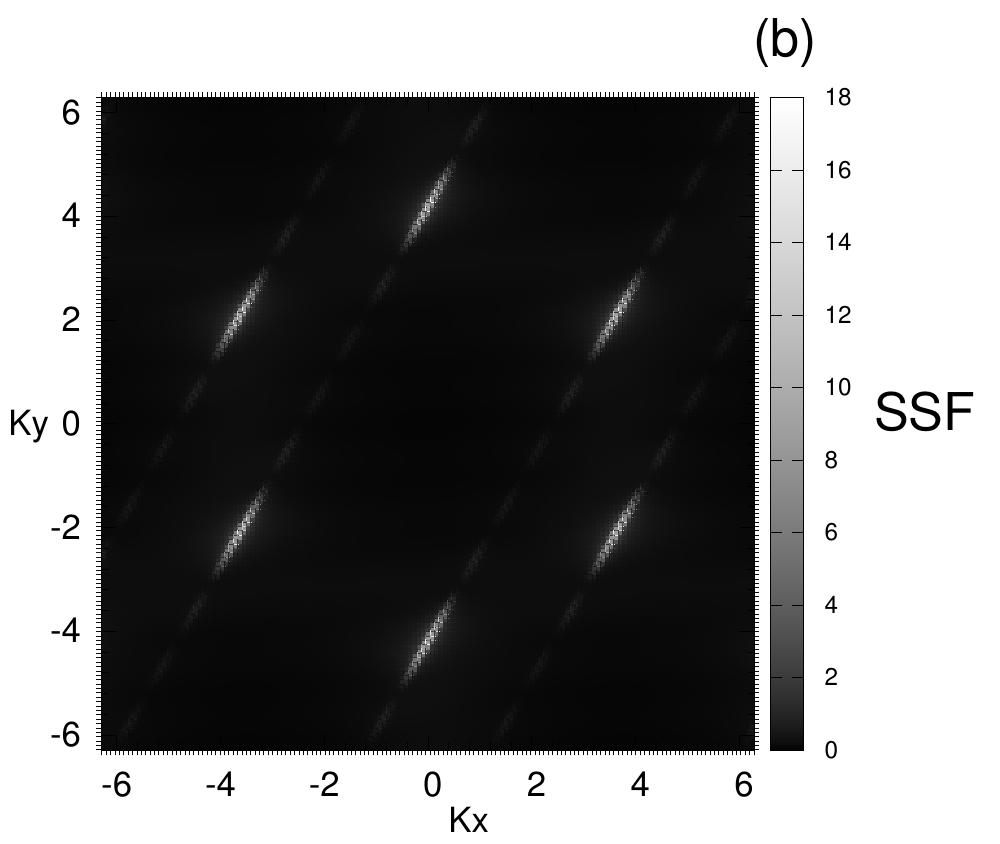}
    \caption{(Color online)
    Lattice visualizations for the iDMRG ground state
    of the LR-TQIM, \eref{eq:HamLR}, on an YC6 structure at $(\alpha,\Gamma)=(3.0,0.3)$ [LR-correlated clock $(0.5,-0.5,0)$-order]. 
    (a) $S^z$-$S^z$ correlation functions for up to 9 legs of the infinite cylinder. The 
    size and the color of the spheres indicate the (long-range) spin-spin correlations 
    in respect to the principal (gray) site, and the thickness and the color of the 
    bonds indicate the strength of the NN correlations. (b) SSF, where we present the
    Bragg-type peaks within the second Brillouin zone of the inverse lattice.
    \label{fig:VisualizationsClock}}
 \end{center}
\end{figure}
%%%%%%%%%%%%%%%%%%%%%%%%%%%%%%%%%%%%%%%%%%%%%%%%%%%%%%%%%%%%%%%%%%%%%%%%%%%

%%%%%%%%%%%%%%%%%%%%%%%%%%%%%%%%%%%%%%%%%%%%%%%%%%%%%%%%%%%%%%%%%%%%%%%%%%%
\begin{figure}
 \begin{center}
    \includegraphics[width=0.99\columnwidth]{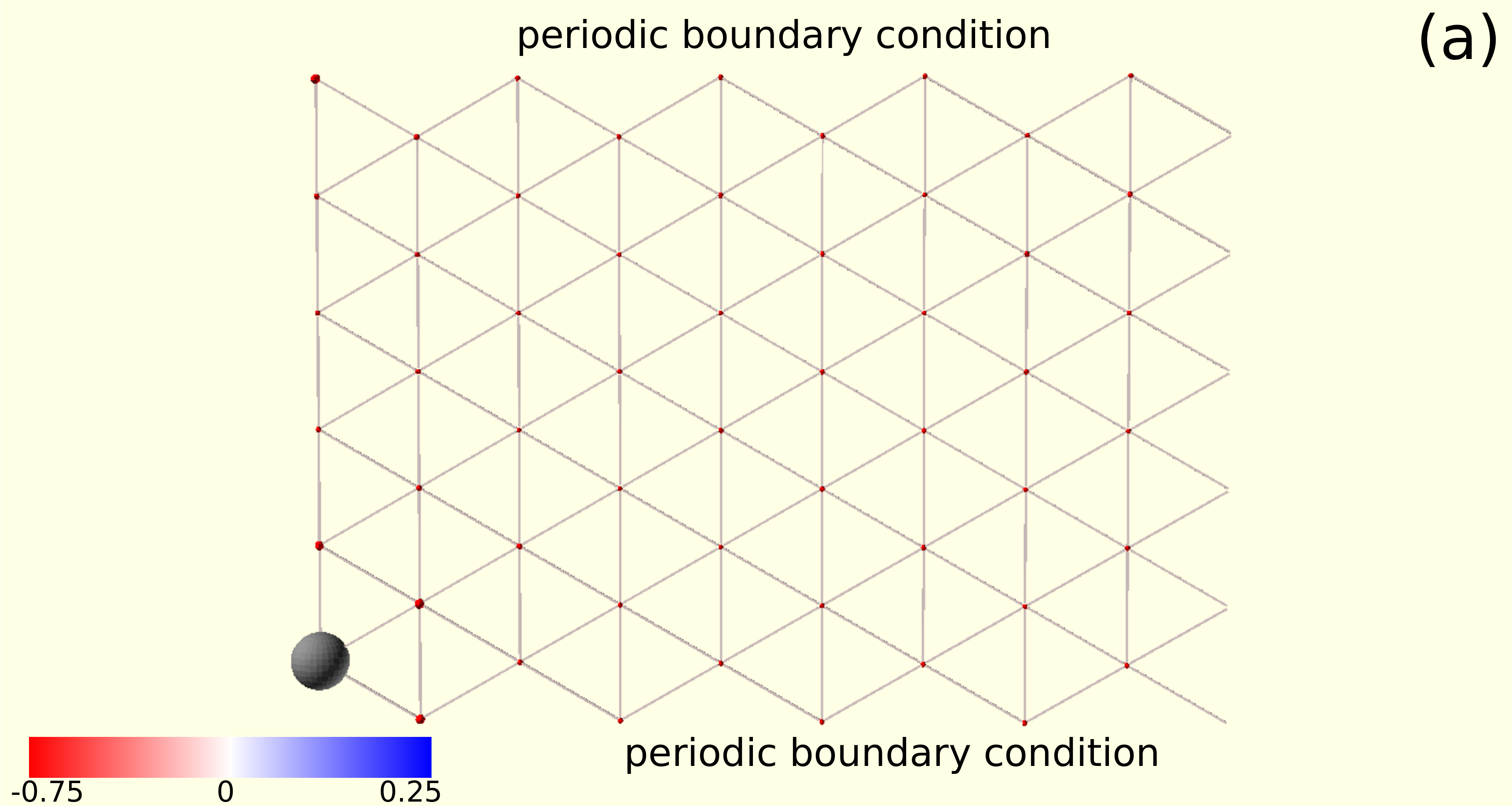}
    \includegraphics[width=0.99\columnwidth]{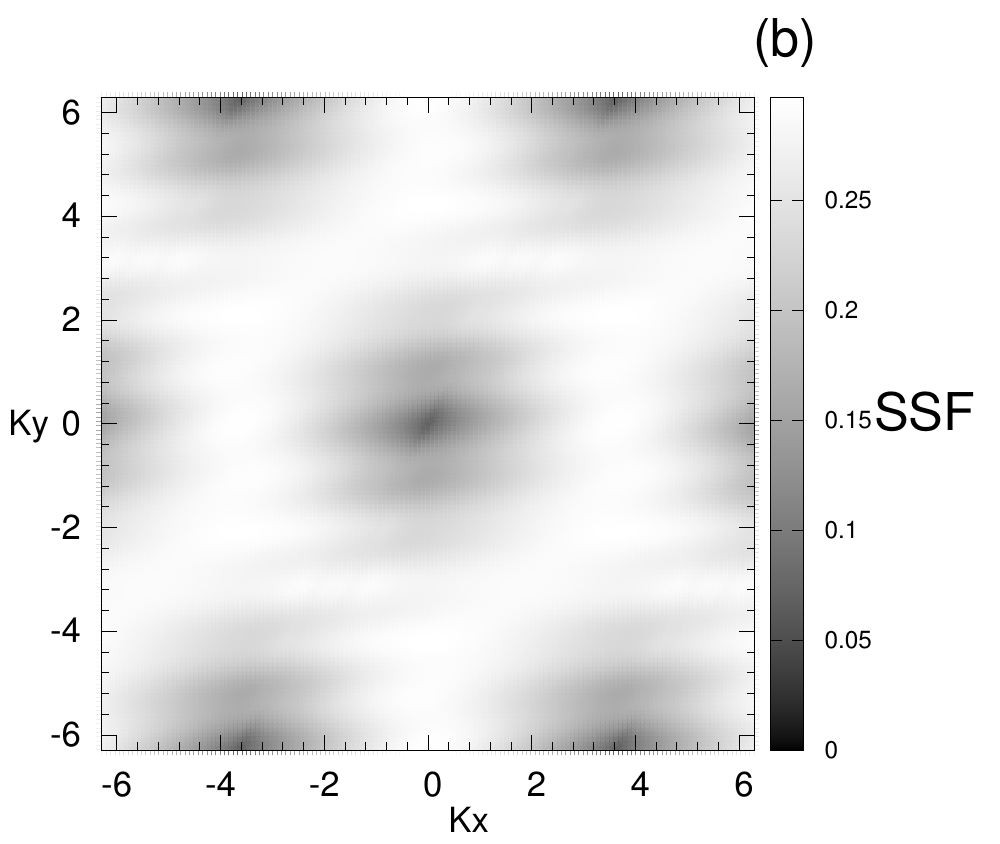}
    \caption{(Color online)
    Lattice visualizations for the iDMRG ground state
    of the LR-TQIM, \eref{eq:HamLR}, on an YC6 structure at $(\alpha,\Gamma)=(1.2,1.5)$ [LR-correlated FM order]. 
    (a) $S^z$-$S^z$ correlation functions for up to 9 legs of the infinite cylinder. The 
    size and the color of the spheres indicate the (long-range) spin-spin correlations 
    in respect to the principal (gray) site, and the thickness and the color of the 
    bonds indicate the strength of the NN correlations. (b) SSF, where there are no
    Bragg-type peaks present within the first and second Brillouin zones of the inverse lattice.
    \label{fig:VisualizationsFM}}
 \end{center}
\end{figure}
%%%%%%%%%%%%%%%%%%%%%%%%%%%%%%%%%%%%%%%%%%%%%%%%%%%%%%%%%%%%%%%%%%%%%%%%%%%

%%% Paragraph 2: detailing the properties of detected ground states for the LR model
We scrutinize the properties of detected ground states of $H_{LR}$ by 
considering some more iDMRG observables in the following subsections.

\subsection{The LR-correlated columnar AFM ordered phase}

%%% Paragraph 1:
We now investigate the properties of the LR-correlated columnar AFM order, which is the `blue' region in \fref{fig:PhaseDiagramLR}.
The ground state is a two-sublattice $Z_2$-symmetry-broken AFM columnar (or stripe) 
order. The phase is columnar in the sense that there exist FM columns (or stripes) spiraling 
the cylinder in the long-direction. The columnar order is two-fold 
degenerate (e.g.~see~Ref.~\onlinecite{Saadatmand17}) on large-width YC-structured triangular lattices as the FM stripes can be aligned either
in $\vektor{a}_{+60^\circ}$ or $\vektor{a}_{-60^\circ}$ directions [see~\fref{fig:TriYC6-generic} -- we note that one can always
set the iDMRG unit-cell size such that the state converges to the arrangement that has 
$\vektor{a}_{+60^\circ}$-aligned FM stripes]. It is noteworthy that columnar order is, in principle, three-fold
degenerate in the true 2D limit~\citep{Lecheminant95}, as the FM stripes can also align in the lattice $Y$-direction; however, 
on YC structures with a large enough width, such iDMRG ground states possess higher energies per site compare to the two 
other alignments. We verified the columnar-ordered nature of spins for this region by observing
the large finite values of $M_2^z(m\rightarrow\infty)$ (vanishing values of 
$M_1^x(m_{\max})$ and $M_3^z(m_{\max})$), [cf.~\fref{fig:PhaseDiagramLR}] and the real-space visualization [projected into a plane]
of calculated $S^z$-$S^z$ correlation functions, as pictured in~\fref{fig:VisualizationsColumnar}(a) for $(\alpha,\Gamma)=(1.2,0.3)$. Furthermore, 
in~\fref{fig:VisualizationsColumnar}(b), we verify the sublattice structure
of the columnar order at $(\alpha,\Gamma)=(1.2,0.3)$ by calculating the static spin structure factor (static SSF) of $S^z$-$S^z$ correlations,
$ \text{SSF}(\vektor{k}, N^{SSF}_{\rm cutoff}) = \frac{1}{N^{SSF}_{\rm cutoff}} \sum_{i,i'}^{N^{SSF}_{\rm cutoff}} \la S^z_{i}
S^z_{i'} \ra e^{i \vektor{k} \cdot (\vektor{r}_{i} - \vektor{r}_{i'})}$,
for large cutoff, $N^{SSF}_{\rm cutoff} \gg 1$, set as the upper limit for site numbers 
(see~\onlinecite{Saadatmand17} for the details of our approach to measure the SSF for iDMRG wavefunctions; 
in particular, here, $\vektor{r}_{i}$ denotes the position vector of a spin,
$S^z_i$, in the \emph{planar} map of the periodic lattice). In \fref{fig:VisualizationsColumnar}(b), the existence of a equilateral parallelogram-shaped 
inverse lattice that surrounds the first Brillouin zone and exhibits %almost $60^{\circ}$ angles and 
four strong Bragg-type peaks, is definitive evidence for the columnar arrangement of spins. 
\fref{fig:VisualizationsColumnar}(b) predicts the wave vector of $\vektor{Q}=(\pm1.86(6),3.12(4))$ for this phase, which is
quite close to the expected vector of $\vektor{Q}_{\text{columnar}}^{\text{theory}} = (\pm\pi/\sqrt 3,\pi) \approx (\pm1.81,3.14)$.

%%% Paragraph 2:
The SR-correlated version of the columnar order was previously observed as the ground state
of the $J_1$-$J_2$ triangular Heisenberg model on the 
YC structures~\citep{Zhu15,Hu15,Saadatmand15,Saadatmand17}, which emerges from continuous symmetry breaking for large positive 
$J_2/J_1$ (considering antiferromagnetic $J_1$). The properties of the columnar order of $H_{LR}$ are virtually
the same as this SR-correlated columnar phase, except, importantly, we discovered that for the former, the LR quantum 
fluctuations in the Hamiltonian leads to \emph{LR correlations}. This 
is evident from the power-law decay of connected correlation functions as shown in \fref{fig:LR-CorrFunc}(a) for $(\alpha,\Gamma)=(1.2,0.3)$.
In addition, the columnar order is LR-entangled due to possessing large correlation lengths
as shown in \fref{fig:LR-CorrLength} for $(\alpha,\Gamma)=(1.2,0.3)$. We note that the saturation of 
correlation lengths (versus $m$) for considered control parameters
are only due to the existence of a finite $n_{\rm cutoff}$ in our LR to exponential-decaying couplings approximation,
\eref{eq:ExpExpansion}, and does not convey any physical meaning. This would be eventually
true for any correlation length curves of types plotted
in \fref{fig:LR-CorrLength}, in case one continues to find $\xi$-values for larger-$m$ ground states. 
Nevertheless, although we have not measured the spin gap directly, the columnar phase has coexistence
of magnetic ordering and power-law correlations due to the LR interactions, and hence we expect that
the spectrum is gapless.
% finite spin gap, due to discrete symmetry breaking and the gapped nature of counterpart two-sublattice AFM ground states~\citep{Vodola16} of one-dimensional $H_{LR}$.
%gapless spectrum due to LR correlations that in turn suggests the coexistence of the discrete-symmetry-broken LR magnetic order and an algebraic paramagnetic state.      
%(in addition, we observed that the columnar ground states possess large finite entanglement spectrum gaps 
%when considering a bipartite cut on the infinite cylinders 
%-- not detailed here).

\subsection{The LR-correlated clock $(0.5,-0.5,0)$-ordered phase}

%%% Paragraph 1:
We now turn to the LR-correlated clock $(0.5,-0.5,0)$-order, shown as the `red' region of \fref{fig:PhaseDiagramLR}.
The ground state is a three-sublattice $Z_2$-symmetry-broken clock order arranged antiferromagnetically on a triangular plaquette
according to $(0.5,-0.5,0)$, which exhibits LR correlations and LR entanglement. The sublattice
properties of the $(0.5,-0.5,0)$-order of $H_{LR}$ are \emph{exactly} the same as the LR-correlated clock order of the NN model, 
\sref{sec:PhaseDiagramNN}, except that the LR correlations are now predicted to be (at least partly) 
induced by LR interactions in $H_{LR}$. We verified the sublattice structure of the 
LR-correlated clock order using the measurement of large finite values of $M_3^z(m\rightarrow\infty)$ (vanishing values of 
$M_1^x(m_{\max})$ and $M_2^z(m_{\max})$) -- see~\fref{fig:PhaseDiagramLR}.  We performed
visualizations of real-space correlations, as shown in 
\fref{fig:VisualizationsClock}(a) for $(\alpha,\Gamma)=(3.0,0.3)$, and calculating the SSF, as shown
in \fref{fig:VisualizationsClock}(b), at the same point. 
In \fref{fig:VisualizationsClock}(b), the existence of a hexagonal-shaped inverse lattice that surrounds the first
Brillouin zone and exhibits six strong Bragg-type peaks, shows that there is a three-sublattice arrangement of the spins. 
\fref{fig:VisualizationsClock}(b) predicts the wave vector of $\vektor{Q}\approx(\pm3.61(5),\pm2.06(6))$ for this phase, which is
quite close to the expected vector of $\vektor{Q}_{\text{clock}}^{\text{theory}} = (\pm2\pi/\sqrt 3,\pm2\pi/3) \approx (\pm3.63,\pm2.09)$. 
Furthermore, we verified the LR-correlated nature of the 
phase by observing power-law decay of connected correlators (at least for short distances) as demonstrated 
for $(\alpha,\Gamma,m)=(4.0,0.3)$ in \fref{fig:LR-CorrFunc}(b). As in \fref{fig:NN-CorrFunc}(a) for the NN model,
in \fref{fig:LR-CorrFunc}(b) [which belongs to a $m=250$ wavefunction], 
it seems that the correlator tails drop exponentially fast; we again argue that this is a 
finite-$m$ phenomenon and for $m\rightarrow\infty$, one would recover an ideal algebraic decay 
(when we decreased the number of states, 
the exponential-drop tail started to appear, always, at shorter distances). Moreover, the ground state is LR-entangled 
due to exhibiting a power-law increase of correlation lengths, as shown in \fref{fig:LR-CorrLength} for $(\alpha,\Gamma)=(3.0,0.3)$,
which goes up to $\xi(m_{max})=O(10)$ per Hamiltonian unit-cell size (however, we reiterate that the correlation lengths 
can still start to saturate for larger $m$). 
At last, we expect the LR-correlated clock $(0.5,-0.5,0)$-order to be gapless % gapped 
due to the same reasoning provided for the gap nature of the LR-correlated columnar ground states.

\subsection{The LR-correlated $x$-polarized FM ordered phase}

%%% Paragraph 1:
Finally, we analyze the LR-correlated $x$-polarized FM order, shown as  the `gray' region of \fref{fig:PhaseDiagramLR}.
The ground state is a ferromagnet with spins exhibiting partial polarizations in spin's $x$-direction and 
paramagnetic in other directions, while possessing LR correlations and LR entanglement. The spin alignment
properties of the FM order of $H_{LR}$ are virtually the same as the FM order of the NN model, 
\sref{sec:PhaseDiagramNN}, except, importantly, the former is LR-correlated. We verified the FM arrangement of the spins 
in the ground state by measuring large finite values of $M_1^x(m\rightarrow\infty)$ 
[i.e.~non-zero net magnetization; also, $M_2^z(m_{\max})$ and $M_3^z(m_{\max})$ are vanishing in this region], cf.~\fref{fig:PhaseDiagramLR},
visualization of real-space correlations, as shown in \fref{fig:VisualizationsFM}(a) for $(\alpha,\Gamma)=(1.2,1.5)$, and calculating the SSF, as shown
in \fref{fig:VisualizationsFM}(b) for the same point. In \fref{fig:VisualizationsFM}(b), 
there are no significant Bragg-type peaks within the first and second Brillouin zones (the SSF 
is featureless in this sense) that verifies the paramagnetic nature of the FM order considering  
$S^z$-$S^z$ correlations. Furthermore, we verified the LR-correlated nature of the 
phase by observing power-law decay of connected correlators as demonstrated 
for $(\alpha,\Gamma)=(1.2,1.5)$ in \fref{fig:LR-CorrFunc}(c). Moreover, the LR-entangled nature of the ground state is clear from the 
power-law increase of correlation lengths, as shown in \fref{fig:LR-CorrLength} for $(\alpha,\Gamma)=(1.2,1.5)$ 
(the correlation lengths are increasing to values as large as $\xi(m_{\rm max})=O(1000)$ per Hamiltonian unit-cell size, and then,
saturate due to existence of a finite $\mathcal{L}_{\text{cutoff}}$).

\section{Conclusion and outlook}
\label{sec:conlusion}

%%% Paragraph 1: summarizing our approach and concluding our findings.
We have exploited the latest developments in iMPS and iDMRG algorithms~\citep{McCulloch08,Crosswhite08,Michel10} to calculate 
fully-quantitative phase diagrams of NN- and LR-interacting triangular Ising models in a transverse field on 6-leg infinite-length cylinders.
%These methods allow us to represent exponential-decaying interactions of a translation-invariant MPS as a 
%Schur-form MPO, and then calculates the expectation values per site using 
%the transfer operator method. 
%To transform exponential-decaying couplings to power-law ones, we employed
%the finite-term sum-of-exponentials approximation presented in \eref{eq:ExpExpansion}. 
The phase diagram of the
NN model contains a LR-correlated clock $(0.5,-0.5,0)$-order, a partially-polarized SR-correlated FM order, and a second-order phase transition
at $\Gamma=0.75(5)$, which agrees relatively well with the results of 
Refs.~\onlinecite{Penson79,MoessnerSondhi01,Isakov03}. More interestingly, for the LR-TQIM, the phase diagram 
hosts a LR-correlated columnar order, a LR-correlated clock $(0.5,-0.5,0)$-order, a LR-correlated $x$-polarized FM order,
and second-order phase transition lines in-between.  Notably, the detected clock order 
is different from the clock order found by recent mean-field/QMC 
results from Ref.~\onlinecite{Humeniuk16}. Our numerical results argue for that
in ladder-type highly-frustrated two-dimensional magnets: the LR quantum 
fluctuations always lead to LR correlations in the ground states. 
% In other words, we established a new type of correlation scaling, incorporated with discrete symmetry breaking, 
% in geometrically-frustrated quantum matter. 
We expect our numerical claims to be justifiable in future ion-trap experiments and can be tested in forthcoming numerical simulations.

%%% Paragraph 2: remaining open questions and possible future works.  
Our work raises several open questions regarding LR interactions in triangular lattices, 
and provides some future research directions.
Our results constitute the first simulation of such systems using iDMRG;
however they are restricted to 6-leg infinite cylinders.  We have shown
that $L_y=6$ is large enough to provide higher than one-dimensional physical phenomena, 
while also being the smallest size that respects the tripartite symmetry
and other requirements.  Next research could investigate the phase diagram of the highly-frustrated $H_{LR}$ 
on larger width cylinders to study the effect of the width on phase stabilization;
this was successfully implemented for the SR-interacting $J_1$-$J_2$ triangular Heisenberg model~\citep{Saadatmand17} on YC8, 10, 12. 

%%% Paragraph 3:
Working towards simulations on larger lattices is also important in the context of 
experimental quantum simulators based on trapped ions, which are achieving increasing numbers 
of spins in their simulation.  The current state-of-the-art is 219 spins on a disk-shaped cluster~\citep{Bohnet16},
with physics that may more closely approximate the true 2D limit rather than the cylinder. The former 
is a limit that iDMRG simulations can describe more accurately as $L_y$ increases. 

%%% Paragraph 4:
The dynamics of such quantum 
simulators is also of interest, often more so than the static ground state properties. 
Time-dependent variation principle~\citep{Haegeman11} and MPO-based~\citep{Zaletel15} algorithms can be 
already used to time-evolve an iMPS subjected to LR couplings; some progress in understanding the dynamics of the 
LR-TQIM on infinite cylinders has been already made by employing another MPO-based time-evolution approach~\citep{HashizumePrep}. 
Further developments of such algorithms
may also open a path for finding finite-temperature states through the imaginary-time simulations. 

%%% Paragraph 5:
Finally, our work highlights several foundational open questions of interest to both quantum information and condensed matter
physicists.  Is there a universal entanglement entropy scaling law for the 
LR-correlated phases in two dimensions? If there is, what are the corrections to the expected
area-law of entropy as found for LR Hamiltonian in one dimension~\citep{Koffel12}? Similar to the significance of the area-law of entropy for
local gapped Hamiltonian, which provides the main reason behind the enormous success of MPS/DMRG for SR interactions, answering this question 
will assist in our collective attempt to fully classify LR-correlated quantum matter.

\begin{acknowledgments}
% will be added at the very end.

  The authors would like to thank Aroon O'Brien for discussions in the early stages of the project. 
  S.~N.~S.~and I.~P.~M.~would also like to thank Tomohiro Hashizume %and Anders Sandvik  
  for useful discussions and suggestions. This research was supported by the 
  Australian Research Council (ARC) Centre of Excellence for Engineered Quantum Systems (EQuS), project number
  CE110001013, and by the ARC Centre of Excellence for Quantum Computation and Communication Technology (CQC2T), 
  project number CE170100012. In addition, %S.~N.~S.~acknowledges the substantial support from Centre for Quantum Dynamics, Griffith University and 
  I.P.M.~acknowledges the support from the ARC Future Fellowships scheme, FT140100625.
   
\end{acknowledgments}

%\bibliographystyle{unsrtnat}

%%%%%%%%%%%%%%%%%%%%%%%%%%%%%%%%%%%%%%%%%%%%%%%%%%%%%%%%%%%%%%%%%%%%%%%%%%%
%%%%%%%%%%%%%%%%%%%%%%%%%%%%%%%%%%%%%%%%%%%%%%%%%%%%%%%%%%%%%%%%%%%%%%%%%%%

%%%%%%%%%%%%%%%%%%%%%%%%%%%%%%%%%%%%%%%%%%%%%%%%%%%%%%%%%%%%%%%%%%%%%%%%%%%
%%%%%%%%%%%%%%%%%%%%%%%%%%%%%%%%%%%%%%%%%%%%%%%%%%%%%%%%%%%%%%%%%%%%%%%%%%%

\cleardoublepage

\end{document}